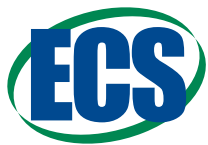

# Temperature-Dependence of the Solid-Electrolyte Interphase Overpotential: Part I. Two Parallel Mechanisms, One Phase Transition

Michael Hess [1,2,z]

[1]*Department of Electrical Engineering and Information Technology, ETH Zurich, Laboratory of Nanoelectronics, 8092 Zurich, Switzerland*
[2]*Battronics AG, β-research division, 8037 Zurich, Switzerland*

It has been shown recently that the overpotential originating from ionic conduction of alkali-ions through the inner dense solid-electrolyte interphase (SEI) is strongly non-linear. An empirical equation was proposed to merge the measured resistances from both galvanostatic cycling (GS) and electrochemical impedance spectroscopy (EIS) at 25°C. Here, this analysis is extended to the full temperature range of batteries from −40°C to +80°C for Li, Na, K and Rb-metal electrodes in carbonate electrolytes. Two different transport mechanisms are found. The first one conducts alkali-ions at all measured temperatures. The second transport mechanism conducts ions for all seven measured Li-ion electrolytes and one out of four Na-ion electrolytes; however, only above a certain critical temperature $T_C$. At $T_C$ a phase transition is observed switching-off the more efficient transport mechanism and leaving only the general ion conduction mechanism. The associated overpotentials increase rapidly below $T_C$ depending on alkali-ion, salt and solvent and become a limiting factor during galvanostatic operation of all Li-ion electrolytes at low temperature. In general, the current analysis merges the SEI resistances measured by EIS ranging from 26 $\Omega cm^2$ for the best Li up to 292 M$\Omega cm^2$ for Rb electrodes to its galvanostatic response over seven orders of magnitude. The determined critical temperatures are between 0–25°C for the tested Li and above 50°C for Na electrolytes.

Extensive research over the last thirty years on Li-ion batteries (LIB) has advanced the field so that LIBs do not only power electronic gadgets anymore as in the 1990's but are also starting to drive new transportation systems like electric bikes, cars and buses. However, in transportation performance at different temperatures is far more relevant especially during recharge.

Recently, it was shown that the overpotential originating from the solid-electrolyte interphase (SEI) is strongly non-linear.[1,2] At ambient temperature, this overpotential is negligible for Li-metal in alkyl-carbonate solvents, however, severe for Na and K-metal electrodes. The resistances from both, electrochemical impedance spectroscopy (EIS) and short galvanostatic charge and discharges (GS) could be fit with the same parameter set by introducing an empirical equation.[1]

However, that publication focused exclusively on standard conditions of ambient temperature and fixed concentrations of solvents and salts commonly used in research.[1] Naturally, the problem needs to be investigated at different conditions including temperature $T$, pressure, and concentration $c$ of various species within the system. While the influence of concentration is rather of interest for patents, temperature dependence is both, crucial for academic understanding and for industrial application.

Very precise measurements of the SEI resistance have been published using EIS. These resistances vary for different Li salts and solvents usually between 40 to 100 $\Omega cm^2$.[1,3–6] In contrast, for Na-metal electrodes these SEI resistances rise already to 230–670 $\Omega cm^2$, while they reach 3500 $\Omega cm^2$ for 0.5 M KPF$_6$ in EC:DMC 1:1.[1] In contrast, galvanostatic cycling of the K-system is possible up to current densities of 56 mAcm$^{-2}$,[1] with as little as 1.5 V total overpotential from both electrodes. If one would assume linear ohmic resistance for the SEI as often applied in modeling,[7] an overpotential of incredible 196 V (3500 × 0.056) would be expected instead of the measured 1.5 V. Thus, SEI resistance is clearly not of ohmic character.

In the 1980's and beginning of 1990's, intense research has been focused on the SEI, also with regard to its overpotential. Many studies concentrated on the system of LiAlCl$_4$ in SOCl$_2$ solvent leading to a very thick SEI layer on the Li-metal electrode due to the highly reducing character of Li and the good oxidant SOCl$_2$.[8–10] Also LiClO$_4$, LiAsF$_6$, and LiBF$_4$ in PC, and γ-butyrolactone[10–13] have been analyzed. For these systems two different models of ionic conduction based on Young's equation[14,15] and space-charge limited current[10,13] have been successfully applied. However, it was not possible to fit the SEI overpotential of Li, Na, and K electrodes at ambient temperature in several different carbonate systems with neither of the two proposed models.[1]

Looking into literature, the space-charge limited model[10] was already disqualified by the same authors in their temperature-dependent data in plot 3 of Ref. 13. They showed a power-law-dependence with the exponent ranging from 2.5 for their 70°C data up to 4.5 at −35°C while they expected a power law dependence of two. While Churikov extended the model to a temperature-dependent exponent to fit the data,[13] one has to mention, that a power law is not expected in electrochemistry as most equations are based on the Euler number (Butler-Volmer, Nernst, Arrhenius, conduction in concentrated electrolytes).[7]

The Young equation seems to be the most prominent and is cited throughout literature from its development in 1961[14] to its proposition for the SEI in 1979[16] and through SEI research and reviews up to today.[6,17] The equation by Young has the form i = prefactor · sinh(zF/RT · a/d · η)[14] with $a$ being the half-jump distance and $d$ the thickness of the film which is intended to describe the hopping of ions through Frenkel and Schottky defects.[14] One can simply challenge the validity of this equation by its limits. If we assume no defects at all, $a$ is 0.5d so ions hop through the whole SEI and the Young equation becomes the Butler-Volmer equation. When we assume infinite many defects, $a$ becomes 0 and the needed overpotential approaches infinite to drive at least a certain current. Thus, the more defects we have the higher the needed overpotential up to infinity for a very defect-rich SEI. This is clearly in contrast to general observations where we observe better hopping and conduction if sites are closer together and seems to be due to the fact that the half-jump distance might not be the correct parameter to use.

Additionally, since $a$ is always smaller than 0.5d, the lowest overpotential we could expect is a Bulter-Volmer equation and as soon as defects are introduced more overpotential is needed. Thus, in LIB research, the SEI overpotential would have been a dominant contribution during galvanostatic experiments as it would have always been higher than any charge-transfer reaction of Butler-Volmer type. Thus, also the Young equation cannot be used to describe ionic transport in the SEI or would need to be modified.

To get more insight into the overpotentials associated with the ionic conduction through the SEI, here, temperature dependence is investigated. The results of Li, Na, K and Rb-metal electrodes are compared to literature values and dissected into their underlying conduction mechanisms.

[z]E-mail: michael@battronics.com

### Experimental

Dry Li-foil (Alfa Aeser, 99.9%), dry Na-rods (Acros, 99.8%), K-cubes in mineral oil (Aldrich, 99.5%) and Rb-metal in ampoules (Strem, 99.9+%) have been cleaned from oxidation layers and used to prepare 13 mm diameter electrodes. Symmetrical alkali vs. alkali cells were prepared in an Argon filled glove box with continuous removal of $O_2$, $H_2O$ and organic volatiles being usually <0.1 ppm. However, K-metal and especially Rb oxidized very quickly forming a purple (K) to blue (Rb) surface layer within a minute and few seconds, respectively, before cell assembly could be finished. Cs electrodes were impossible to prepare in the glove box due to its low melting point of 28.4°C.

Ethylene carbonate EC (Aldrich, anhydrous 99%), propylene carbonate PC (Aldrich, anhydrous 99.7%), and dimethyl-carbonate DMC (Acros, extradry 99+%) were additionally dried over 4Å molecular sieves for at least six weeks after which 16ppm of trace water was still present measured by Karl-Fischer-Titration. Electrolytes for Li were purchased in prepared state being 1 M $LiPF_6$ in EC:DMC 1:1 wt (Novolyte/BASF).

The salts $LiPF_6$ (Strem Chemicals, 99.9+%), $LiClO_4$ (Aldrich, ampoule 99.99%), Li-bis(oxalato)borate LiBOB (Aldrich), Li-bistri fluoromethanesulfonimidate LiTFSI (Aldrich, 99.95%), $NaPF_6$ (Alfa Aesar, 99+%), $NaClO_4$ (Acros, 99+%), $KPF_6$ (Strem Chemicals, 99.5%), $KClO_4$ (Acros, 99%) and $RbClO_4$ (Strem, 99.9%) were vacuum dried at 25°C for one day before use. The solvents were prepared in weight equivalent mixtures. The salt was added based on the calculated density of the pure solvent mixture without the salt leading to a systematic error of circa 3–4% lower molarity than 1 M (see Suppl. Note 1).

Whatman glass microfiber filters (GE Healthcare, GF/D 1823–257) were heated inside the glove box to 400°C to remove adsorbed water and are used due to their very high porosity of circa 86–88% after compression to circa 430 μm at p = 50 $Ncm^{-2}$ in the coin type cells made from titanium (please note error in reported porosity and thickness of GF separator in Ref. 1). Galvanostatic cycling (GS) and electrochemical impedance spectroscopy (EIS) were performed with Biologic VMP3 and MPG2 cyclers from temperatures ranging from −40 to +80°C for GS and −40 to 25°C for EIS. The maximum allowed overpotential of the cell was 4.5 V to avoid SEI oxidation at the dissolving alkali-metal electrode. More details can be found in Suppl. Note 1.

Experimental limitations are given by solubility limits of Na/K-$PF_6$ in EC:DMC 1:1wt[18] and K/Rb-$ClO_4$ in both EC:DMC and PC. Usually, the maximum solubility of a particular salt in a solvent at 25°C was not prepared to avoid early salt precipitation at lower temperatures. Additionally, the melting point of alkali-metals of $T_m = 63.5°C$ for K and $T_m = 39.5°C$ for Rb limited the upper operation temperature for K/Rb cells. The lower operation temperature of EC:DMC 1:1wt mixtures was also limited where the electrolytes operated last at −20°C for 1 M $LiPF_6$, −30°C for 1 M $LiClO_4$, 1 M LiTFSI and 0.5 M $NaPF_6$, and only until −10°C for 1 M $NaClO_4$, 0.5 M $KPF_6$ and 0.05 M $KClO_4$ all in EC:DMC 1:1wt, while all electrolytes containing PC operated nicely until −40°C due to its low $T_m$ of −48.8°C.[19] These are the hard bounds while further soft limitations will be given in the Results section.

### Results

The measured GS and EIS data are analyzed based on the three transport processes of charge transfer reactions, ionic conduction in the electrolyte and ionic conduction within the inner dense SEI. These three transport processes will be modeled by the Butler-Volmer equation (Eq. 1), Ohms law (Eq. 2) and the empirical equation proposed earlier (Eq. 3).[1,2] While Butler-Volmer and Ohms law are commonly used equations the third empirical equation by Hess[1] does not normalize the overpotential over 25.6 mV at 25°C but introduces a muting factor H in the exponent, thus, making the saturation potential variable. The following three equations are used for the current analysis:

$$j_{ButVol} = j_{0,BV}(e^{0.5 \frac{zF}{RT} \eta_{BV}} - e^{-0.5 \frac{zF}{RT} \eta_{BV}}) \quad [1]$$

$$j_{Ohm} = \eta_{Ohm}/AR_{Ohm} \quad \text{where} \quad R_{Ohm} = R_{elyte} \cdot \tau_{sep}/\varepsilon_{sep} \quad [2]$$

$$j_{Hess} = j_{0,H}(e^{0.5 \frac{zF}{RT} \eta_H \cdot H} - e^{-0.5 \frac{zF}{RT} \eta_H \cdot H}) \quad [3]$$

with the parameters current densities $j$, exchange current densities $j_{0,BV}$ and $j_{0,H}$, overpotentials $\eta_x$, ohmic resistance $R_{Ohm}$, the muting factor $H$, Faraday constant $F$, gas constant $R$, and the temperature $T$. One needs to assume that all three processes are in series so that $j_{BV} = j_{Ohm} = j_{Hess}$ meaning the overpotentials are additive. This assumption is valid as long as all current passes through all processes and all processes are sufficiently homogeneous over the alkali electrode to average them macroscopically. Additionally, all kinetics are neglected, e.g. double layer capacitance variations, SEI reforming processes and the geometric surface area is used neglecting dendrite growth and surface roughness.

To allow comparison between EIS and GS cycling, the limit of the equations of Butler-Volmer, Ohm and Hess for very small excitations can be derived as follows:[1]

$$R_{BV}\left(\eta_{BV} << \frac{RT}{F}\right) = \frac{\eta_{BV}}{jA} = \frac{RT}{FA} \frac{1}{j_{0,BV}} \quad [4]$$

$$R_{Ohm} = \frac{\eta_{Ohm}}{jA} = \frac{\tau_{sep}}{\varepsilon_{sep}} R_{elyte} = \frac{\tau_{sep}}{\varepsilon_{sep}} \frac{A}{\kappa l} \quad [5]$$

$$R_H\left(\eta_H << \frac{RT}{FH}\right) = \frac{\eta_H}{jA} = \frac{RT}{FA} \frac{1}{j_{0,H} H} \quad [6]$$

Fitting of galvanostatic data is based on Matlab's Levenberg-Marquard method in the function "fit" by Mathworks. For the three transport equations in Eqs. 1 to 3, the four parameters $j_{0,BV}$, $R_{Ohm}$, $H$ and $j_{0,H}$ are fit as constants for a given current-density to voltage plot. Thus, the system is over-determined and cannot be solved analytically but is fit to minimize a weighted least-square cost function. The sum of the three overpotentials $\eta_x$ of the three processes at each given current density $j = j_{BV} = j_{Ohm} = j_{Hess}$ are compared to the experimental overpotential and the difference is minimized. Detailed information of the fitting procedure and the error bars can be found in SI notes 3, 5 and 6.

***Error analysis and sensitivity.***—It is important to note that the equations of Butler-Volmer and Hess (Eqs. 1 and 3) and their limit for small excitations (Eqs. 4 and 6) are the same if the muting factor $H$ approaches the value of 1. Thus, the parameters $j_{0,BV}$ and $j_{0,H}$ of both equations can become interchangeable and can lead to errors in the parameter estimation in case of values of $H$ around unity.

Additionally, the analysis of alkali-metal electrodes is usually influenced by several different mechanisms that change the parameters of the underlying reaction and transport mechanisms. All alkali-metal electrodes Li, Na, K and Rb undergo dendrite formation during metal deposition while they show no significant pitting but rather a homogeneous dissolution on the counter electrode during galvanostatic cycling (e.g. Suppl. Videos of Ref. 20). This leads to surface area increases and most probably a thinner newly formed SEI layer on the deposition electrode only. Additionally, symmetric cells of these alkali metals cannot account for asymmetries in both the charge transfer reaction and ionic conduction within the SEI. Thus, using an asymmetry coefficient α = 0.5 of the charge transfer reaction and SEI overpotential and dividing the total overpotential of symmetric cells simply by a factor of two to account for both electrodes is a very questionable and probably wrong assumption. However, no further information on neither the asymmetry parameters nor the real active surface area can be determined precisely without the help of e.g. reference electrodes.

To minimize the effect of surface area increase, only short galvanostatic pulses of 7 mAhcm$^{-2}$ have been applied depositing circa 3.4 μm (0.18 mgcm$^{-2}$) in the case of Li. Each high current density charge/discharge was followed by a low current density discharge/charge, respectively, to smooth out any formed dendrites. However, the newly built SEI on the freshly deposited alkali electrode will contribute to the systematic error of this analysis as it might be thinner and less crystalline. Thus, all following evaluations in this manuscript should be seen in the light that alkali-metal electrodes behave asymmetric, but here, need to be evaluated as if deposition and dissolution are symmetric.

In contrast to the previous publication where at least five nominally equal cells have been tested for each configuration to allow statistics,[1] here, usually only one sample was tested due to the limitations of the climate chamber. Only samples at important temperatures have been repeated up to a maximum of four times to get certain statistics.

The second challenge is the sensitivity of the three transport processes and their individual contribution to both the GS and EIS experiments. In general, the ohmic resistance of the electrolyte can be well extracted from both the GS and EIS data as it is very sensitive to high current densities during GS and the real axis intercept at infinite frequency during EIS. One just needs to note that the transference number is 0.4–0.56 for the GS case depending on the separator[21,22] and 1 in the case of EIS as both anions and cations conduct the current in the limit of infinite frequency.

The charge transfer reaction modeled by the Butler-Volmer equation is the least sensitive parameter. During GS cycling $j_{0,BV}$ can be mainly fitted to high current density data where both dendrite growth and surface area increase contribute the most. Additionally, the usually high exchange current densities of metals and here also alkali-metals make this process rather not limiting, so not very sensitive to small changes. Also the resulting EIS resistances for small excitations in Eq. 4 show that the charge-transfer resistance is rather small compared to the SEI resistance and cannot be extracted from EIS data.[1] Thus, only the GS data at high current density can be used; however, will result in large error bars as shown later. This insensitivity could explain why the exchange current density $j_{0,BV}$ varies over several orders of magnitude in literature (e.g. Table 1 in Ref. 23) for similar electrolytes.

The third process concerns the ionic conductivity of the SEI. During GS cycling this process is sensitive giving the smallest error bars for $j_{0,H}$ and medium error bars for $H$ as they can be determined at very low and medium current density, respectively. Also during EIS the combination of $j_{0,H}$ and $H$ in Eq. 6 is very sensitive as it represents the first semi-circle in the Nyquist plots. However, the individual parameters can only be separated during GS experiments as shown earlier.[1]

*Analysis procedure.*—Examples of short galvanostatic cycling experiments at various current densities are displayed in Figure 1 for 1 M LiPF$_6$ in EC:DMC at −10, 30 and 70°C. At −10°C one can see salt depletion at very high current density of 56 mAcm$^{-2}$ which is double the calculated limiting current density of the system.[1] To guarantee similarity of the different electrodes first two high current density activation discharges have been performed to exchange the native passivation layer on alkali metals.

An interesting feature is the overshoot in the beginning of all current densities at −10°C, all medium to high current densities at 30°C and no such overshooting at 70°C. The 70°C experiment is the actual behavior one expects from the concentration gradient built-up of anions and cations in liquid electrolytes. To analyze the data, two different overpotentials have been used being the initial overpotential right at the beginning which is either the initial maximum as shown at −10°C or the data point near 0.005 mAcm$^{-2}$ at 70°C to account for the double layer formation and other processes with small time constants. The second evaluated overpotential is the one at steady state being usually at the end of the short cycling experiment at 0.07 mAcm$^{-2}$

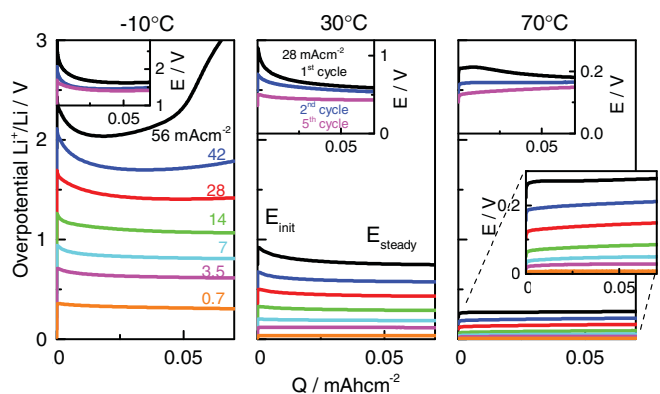

**Figure 1.** Representative example of dependence of galvanostatic overpotentials on temperature and current density: for symmetric cells of Li vs. Li in 1 M LiPF$_6$ EC:DMC 1:1wt for −10, 30 and 70°C, same scale applied to show significant temperature-dependence; insets show first two activation cycles before cycling experiments performed; two different overpotentials used for evaluation: initial potential and the steady-state potential usually at the end of the short cycling experiment.

or at the minimum of the total overpotential if electrolyte depletion occurs.

While the initial overpotential represents a nominally unaltered static SEI layer on both electrodes, the steady state overpotential is believed to represent the SEI during continuous reforming. So the initial overpotential represents an upper bound for the estimated parameters while the steady state overpotential represents a lower bound because the SEI might be thinner.

These overpotentials are extracted for sixteen different electrolyte systems of the four different alkali metals in the range from −40 to +80°C. Figure 2 depicts the three cases of Li/Na/K-PF$_6$ in EC:DMC 1:1 as representatives. The initial overpotentials from Figure 1 have been plotted as markers with the fit shown as a line. The individual contributions are also depicted from the fit to the Butler-Volmer equation, Ohm's law and Hess equation. Usually the overpotential of each individual process decreases with increasing temperature as expected from Arrhenius type of processes. However, at high temperatures of >70°C for Li, >50°C for Na and >30°C for K, the ohmic overpotential in the electrolyte increases again with temperature which will be discussed further in Part II of this series. However, of outmost interest is the collapse of the Hessian overpotentials of Li-metal above 10°C (greenish blue) and for Na-metal above 60°C (red) to negligible small overpotentials. In contrast from −20 to 0°C for Li, −30°C to 50°C for Na, and at all measured temperatures of −10 to 50°C for K, the overpotential according to Hess in Eq. 3 contributes significantly to the total overpotential of these symmetric cells. All sixteen different electrolyte systems are shown in Suppl. Figure S1-S4 for Li, Na, K, and Rb, respectively. It has to be mentioned that the overpotentials based on Eq. 3 for K is already 300–800 mV per electrode and in the case of 0.05 M RbClO$_4$ in PC, this overpotential increases to 1–4 V per Rb-metal electrode. These overpotentials exclude both alkali-metals from real applications besides their significant safety issues as only very little current density can be applied on Rb electrodes until the cutoff potential of 4.5 V is reached.

For comparison, also EIS has been applied from −40°C for PC-based electrolytes or down to the crystallization temperature for EC:DMC electrolytes up to 25°C. These standard Nyquist plots and log-log plots are shown in Figure 3. It is important to note that the ohmic resistance from electrolyte increases by one order of magnitude while the SEI resistance changes between two to four orders of magnitude with temperature.

Evaluation of the first semi-circle was performed based on an equivalent circuits of either $R_1+R_2||Q_2$ or $R_1+Q_2||(R_2+R_3)||Q_3)$ where $R$ and $Q$ represent ohmic resistances and constant phase elements. Usually the most simple circuit is used; however, sometimes a small

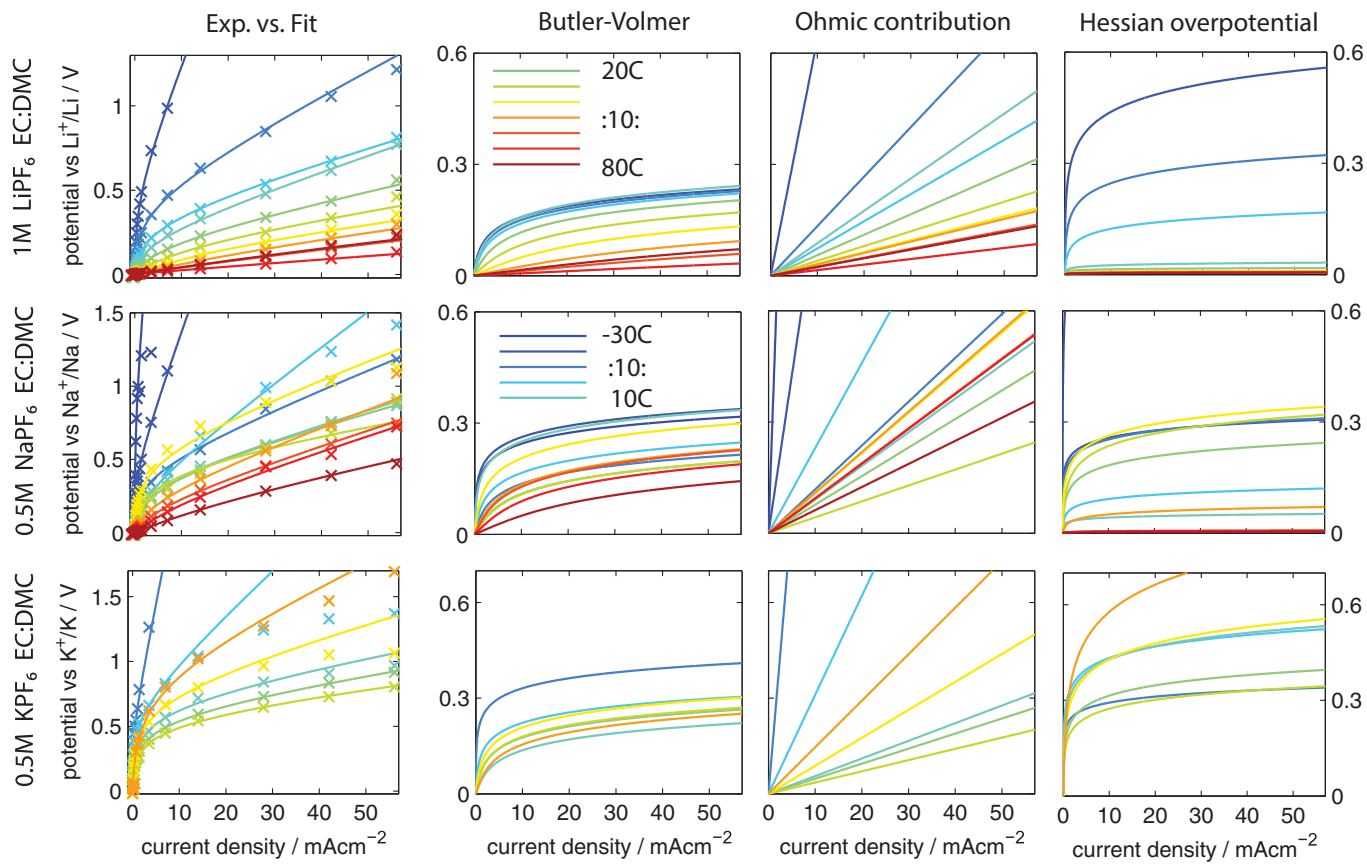

**Figure 2.** Evolution and fitting of initial overpotentials and single contributions for alkali-metal electrodes with temperature: for the three different examples of Li/Na/K-PF$_6$ in EC:DMC 1:1wt soaked in a Whatman glassfiber separator ranging from −20 to 80°C for 1 M LiPF$_6$, −30 to 80 °C for 0.5 M NaPF$_6$, −10 to 50°C for 0.5 M KPF$_6$; while Li-metal tests work well, Na and K-metal are less good in their trends and the ohmic resistance increases at high temperatures against the Arrhenius trend; in general) overpotentials increase from Li over Na to K for each specific temperature where the contribution from charge transfer are in the same range for the three alkali metals except that 1 M LiPF$_6$ has slightly higher exchange current densities than its 0.5 M Na and K counterparts; the ohmic potential drop is higher for 0.5 M electrolytes as expected, however, the main difference comes from the SEI contribution which is negligible for Li in the range of 10–80°C and Na from 60–80°C but very dominant for Li below 0°C, Na below 50°C and for K at all measured temperatures.

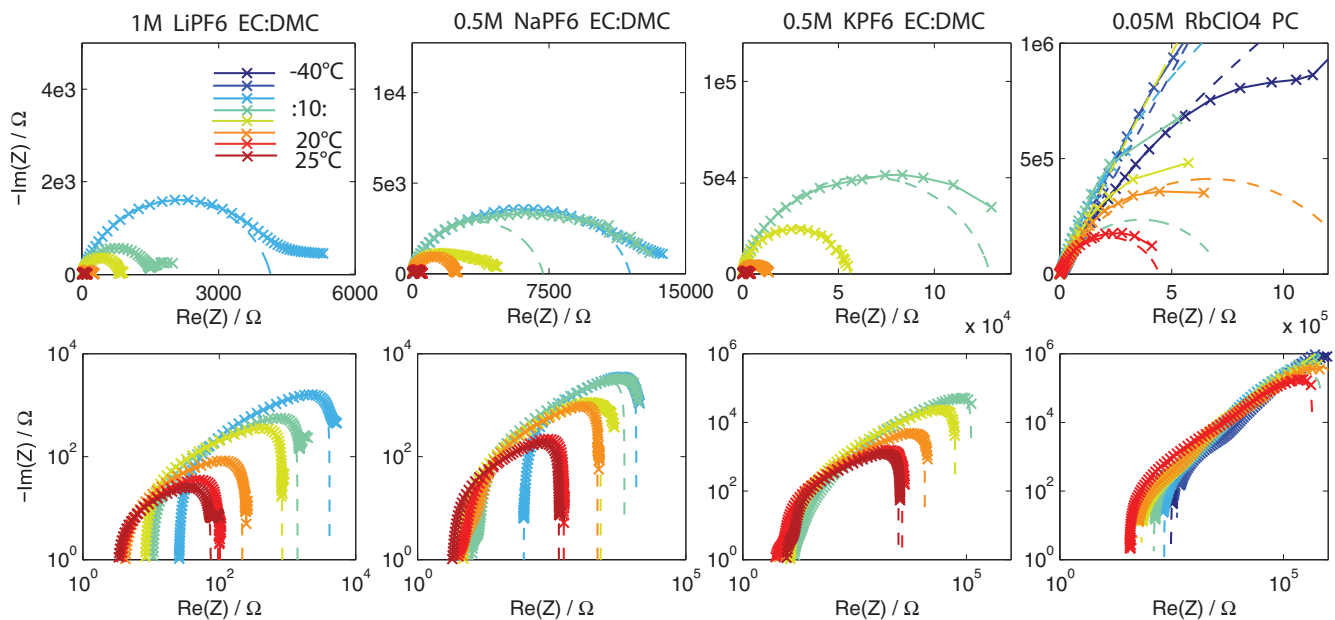

**Figure 3.** Evolution of electrochemical impedance spectroscopy for alkali-metal electrodes with temperature: for four different samples of Li/Na/K-PF$_6$ in EC:DMC 1:1wt and RbClO$_4$ in PC soaked in a Whatman glassfiber separator ranging from −20 to 25°C for 1 M LiPF$_6$ and 0.5 M NaPF$_6$, −10 to 25°C for 0.5 M KPF$_6$, and −40 to 25°C for 0.05 M RbClO$_4$ in PC; 1$^{st}$ row) standard Nyquist plots showing semi-circle evolution, 2$^{nd}$ row) log-log plot to illustrate significant resistance increases over orders of magnitude with decreasing temperature; markers show experimental data at 10 mV excitation from 500kHz to 0.01Hz and dashed line represents fit with equivalent circuits.

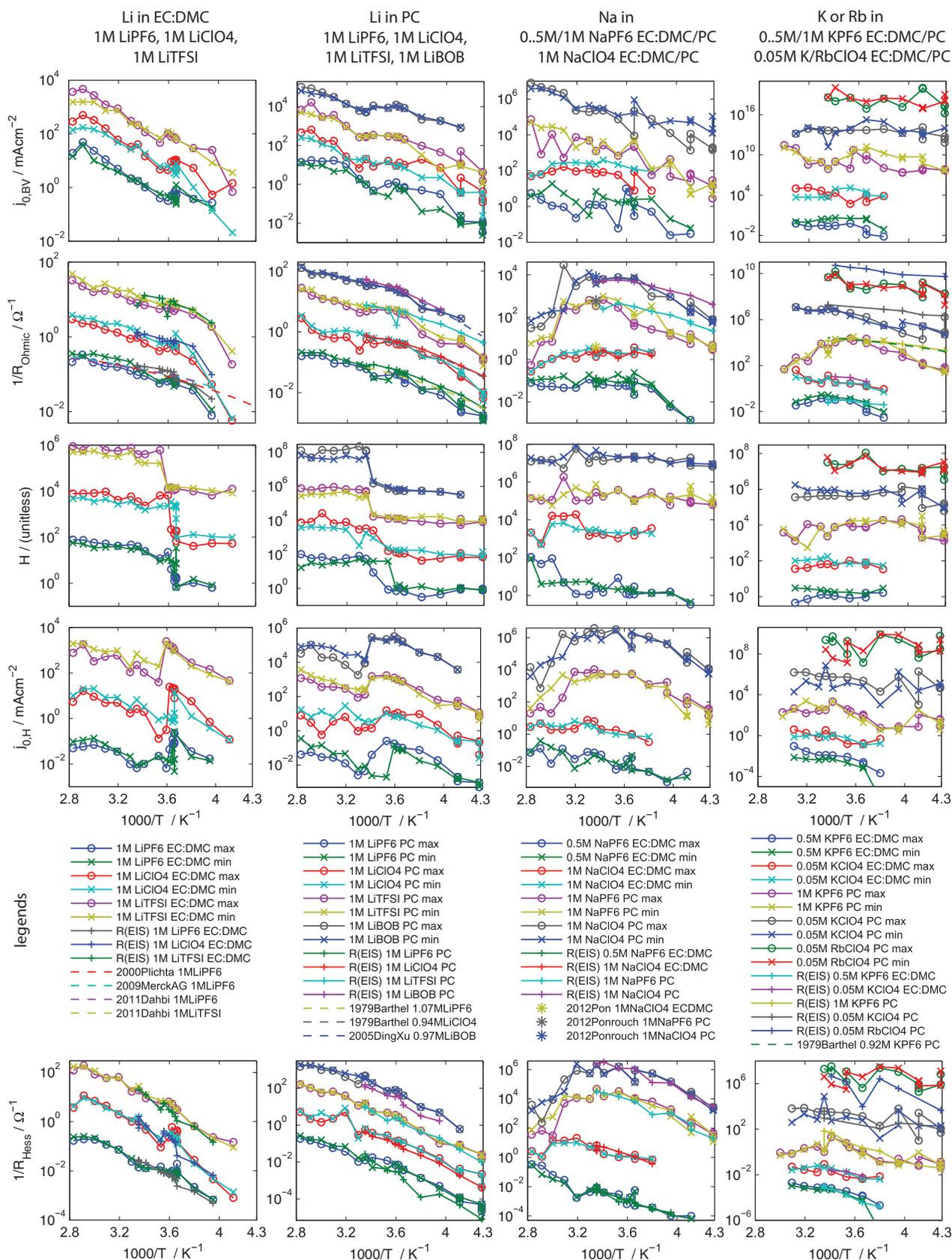

**Figure 4.** Parameter set of best fit of symmetrical Li, Na, K and Rb cells with three processes Butler-Volmer, Ohm and Hess: for sixteen different systems of column: 1) Li in EC:DMC 2) Li in PC, 3) Na and 4) K and Rb with row 1) exchange current density $j_{0,BV}$, 2) ohmic conductance of GS and EIS experiments, 3) muting factor $H$, 4) exchange current density $j_{0,H}$ which are all four independent variables while row 6) plots the derived Hessian resistance combining $H$ and $i_{0,H}$ in Eq. 6 showing the match with the SEI resistance from EIS experiments; electrolyte conductivity data[25–30] has been added in row 2) to show the match.

pre-circle at high frequency needs to be included to fit the standard high-resistance semi-circle. No data at low frequency was evaluated as unphysical high surface areas and small exponents of the constant phase element of values between 0.5 and 0.7 indicate severe issues with either the data or the assumptions of using *RQ* elements as discussed in Ref. 1. Thus, only the electrolyte resistance and the first semi-circle in EIS have been used in this study.

The fit with equivalent circuits is shown as dashed lines in Figure 3. The observed resistances vary between circa 40 Ω for the case of 1 M LiPF$_6$ EC:DMC at 25°C up to 2.2 · 10$^8$ Ω for 0.05 M RbClO$_4$ in PC at −40°C for a geometric electrode area of 1.33 cm$^2$. The lowest resistances of the first semi-circle are found for 1 M LiClO$_4$ and 1 M LiTFSI in EC:DMC with circa 19 Ω at 25°C which is in line with literature[24] and shown in Suppl. Fig. S5, S6. While the SEI resistances for Li electrolytes are still in a range where operation might be possible, the 220 MΩ for Rb would simply represent a heat source if this resistance would be linear throughout the different current densities.

*Extracted parameters of transport processes.*—Figure 4 shows the parameter trends of the exchange current density of Butler-Volmer $j_{0,BV}$ and Hess $j_{0,H}$, the inverse ohmic resistance reflecting the conductivity $1/R_{Ohm}$ and the muting factor $H$ as a function of $1/T$ in Arrhenius plots. While these four parameters are individual variables, the derived value of the Hessian resistance in Eq. 6 is displayed in the last row of Figure 4. Additionally, the electrolyte resistance from EIS corrected for the transference number $t_{Li} = 0.56$[22] and SEI resistance extracted from the fit to the first semi-circle in the EIS experiment have been plotted in the $1/R_{Ohm}$ and $1/R_{Hess}$ plots showing the match for all four alkali-metal electrodes.

The lines in Figure 4 have been arbitrarily shifted to show similarities and respective trends. Each electrolyte system is plotted individually in Suppl. Figure S7-S10 for better visibility. The initial overpotential for each electrolyte, current density and temperature is displayed with circles while the steady state overpotential has x markers reflecting both the maximum and minimum parameter estimate from galvanostatic cycling. The different systems have been sorted for similar behavior as Li salts in EC:DMC, Li salts in PC, Na salts and last K/Rb salts.

While the exchange current density $j_{0,BV}$ shows small fluctuations and typical Arrhenius type of behavior for Li, this exchange current density has enormous errors up to two orders of magnitude in the case of Na, K and Rb metal due to increasing reactivity in the order of Na, K, Rb. In contrast, the ohmic resistance shows small errors and continuous trends. For several electrolyte systems, the measured conductivity for the pure electrolyte could be added in the $1/R_{Ohm}$ plots when adapted for the properties of the glassfiber separator and transference number. This was the case for LiPF$_6$ and LiTFSI in EC:DMC;[25–27] LiPF$_6$, LiClO$_4$ and KPF$_6$ in PC;[28] LiBOB in PC[29] and the conductivity of three Na systems but at 25°C only.[30] These eight electrolyte conductivities match the extracted ohmic resistance. Only, the electrolyte resistance from EIS adapted already by the transference number is in all cases smaller than the one determined from galvanostatic experiments, however, only within a certain error. In general, the electrolyte resistance is the only parameter here, which can be compared to literature values as very precise conductivity meters exist and the conductivities have been tabulated.[25–29]

However, the main feature is the muting factor $H$ and the exchange current density $j_{0,H}$ of the SEI resistance. For all seven Li systems a jump of both $H$ and $j_{0,H}$ is observed with a change of each parameter of one to two orders of magnitude. This jump is sudden and seems to be similar to a phase transition. The critical temperatures are individual for each electrolyte system and seem to be independent from the crystallization temperature of e.g. a 1 M salt in the EC:DMC phase diagram.[31] They are also observed for all four PC-based Li electrolytes. Furthermore, this sudden jump is detected for 0.5 M NaPF$_6$ in EC:DMC; however, at much higher $T_C$ between 50–60°C. No such phase transition is observed for neither of the other three Na systems nor K or Rb. While $j_{0,BV}$, $1/R_{Ohm}$ and $j_{0,H}$ show typical Arrhenius type of behavior, the muting factor $H$ seems to be rather independent of temperature, except at the phase transition near $T_C$.

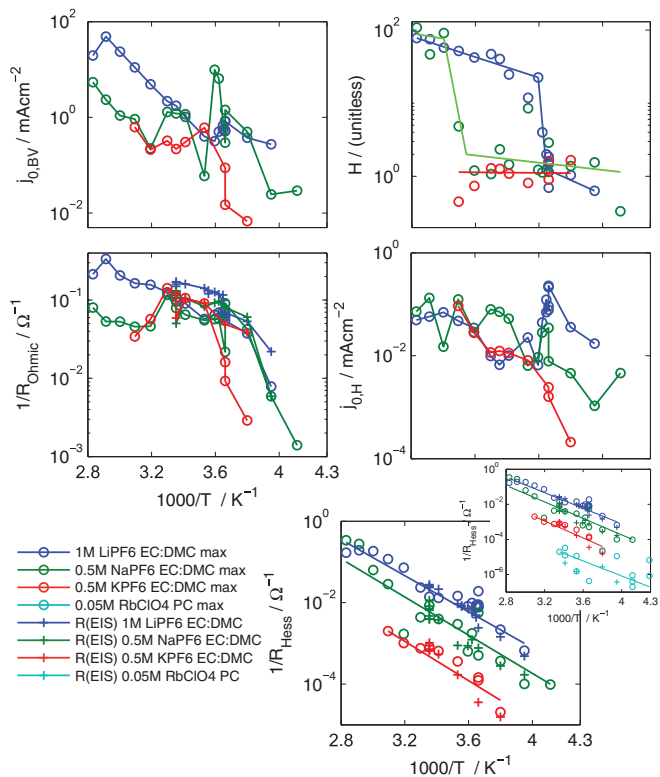

**Figure 5.** Comparison of alkali-PF$_6$ EC:DMC: plots of $j_{0,BV}$, $1/R_{Ohm}$, $H$ and $i_{0,H}$ vs. $T^{-1}$ show an order of magnitude higher exchange current and ohmic resistance for 1 M LiPF$_6$ compared to 0.5 M Na/KPF$_6$; in contrast $H$ seems to be relatively similar for all three electrolytes below $T_C$; however, both Li and Na indicate a phase transition toward high $H$ meaning small SEI overpotentials; also $i_{0,H}$ seems to be rather similar for the three electrolytes within a very high error bar shown from the fluctuation; merging of $H$ and $i_{0,H}$ gives the resistance $R_{Hess}$ which fits for all four alkali metals over seven orders of magnitude to the resistance of the first semi-circle of the EIS experiments.

The data here shows a dilemma for industrial applications. Usually one adds PC to EC:co-solvent mixtures to decrease the crystallization temperature of electrolytes and guarantee a good low temperature operation. However, here all three different 1 M salts of LiPF$_6$, LiClO$_4$ and LiTFSI have circa 10–15°C higher $T_C$ of the SEI resistance in pure PC than in their EC:DMC counter parts. Thus, while one improves ionic conductivity in the electrolyte, one might decrease ionic conductivity in the SEI. The behavior for ternary solvent system is missing here; however, this important aspect should be mentioned at least.

*Comparison of alkali-metal parameters.*—Figure 5 shows the Arrhenius plots of Figure 4 for the cases of 1 M Li-, 0.5 M Na- and 0.5 M KPF$_6$ in EC:DMC 1:1. While the error bar is significant for Na and K, trends can still be extracted as the parameters change over orders of magnitude themselves. While the exchange current density $j_{0,BV}$ and $1/R_{Ohm}$ are smaller for the 0.5 M Na and K electrolytes, as expected, the main difference can be seen in the $H$ parameter which has a sudden jump at certain $T_C$ for both Li and Na, however, circa 50°C apart. It is important to note that the $H$ parameter is very similar for $T<T_C$ for Li/Na and for K at all temperatures. Also above $T_C$, Li and Na seem to have a similar $H$ parameter. Thus, there might be a common transport phenomenon associated with these alkali metals for a certain temperature range. Also $j_{0,H}$ is similar for these different alkali metals except at low temperature below circa 5°C.

The derived Hessian resistance calculated from $j_{0,H}$ and $H$ in Eq. 6 is plotted in the lowest subfigure of Figure 5. One can clearly observe straight lines for Li, Na, K and also Rb in the inset. Despite $H$ and $j_{0,H}$ jumping over two orders of magnitude at $T_C$, they compensate each other and will be invisible during EIS experiments. And indeed, in the last row of Figure 4, both the Hessian resistance calculated from Eq. 6 and the SEI resistance determined from EIS in Figure 3 match and only straight lines with small fluctuations in the regime near $T_C$ can be observed.

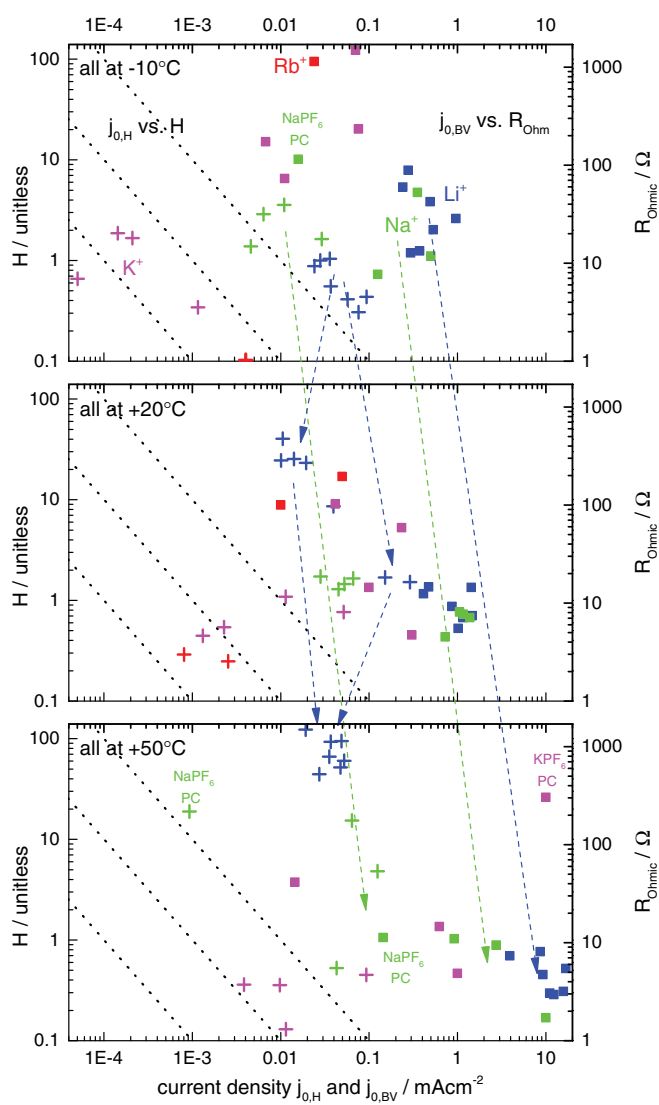

**Figure 6.** Parameter clustering and evolution for the three processes: plots at −10, +20 and +50°C for all sixteen measured electrolyte systems for Li (blue), Na (green), K (magenta) and Rb (red) with markers for $j_{0,H}$ vs $H$ (plus) and $j_{0,BV}$ vs. $R_{Ohm}$ (square); black dotted lines reflect equi-resistance lines for $R_{Hess}$ which is usually measured by EIS, dashed blue and green arrows indicate parameter evolution where the two Li electrolytes LiTFSI and LiBOB in PC have their critical temperature $T_C$ between 20–25°C so splitting of parameter cluster at 20°C, no arrows for K and Rb due to very different salt molarity.

To show the similarity of the four fitting parameters for the three overpotentials, the parameters have been plotted in Figure 6 for −10, +20 and +50°C to maximize the amount of data points. On the left, $H$ is plotted vs. $j_{0,H}$ with equi-resistance lines added to show similar resistance lines from Eq. 6. On the right axis, $R_{Ohm}$ is plotted vs. $j_{0,BV}$ which do not have a physical dependence or correlation. One can observe that the charge transfer reaction is very similar for all seven Li electrolyte systems and three out of four Na systems. This is consistent throughout the different temperatures. K and Rb are not compared as they have very different molarity ranging from 0.05 M to 1 M and enormous error bars due to their high reactivity and dendrite growth rate. These error bars are shown in Suppl. Figure S11 for each individual alkali-metal parameter at the three temperatures. Based on these error bars, one can confirm that $j_{0,BV}$ has the lowest sensitivity in GS experiments due to very high exchange current density for metals in general and therefore non-limiting character. In contrast, $j_{0,H}$ and $H$ have relative small error bars and are usually the most sensitive during GS and EIS measurements, while $1/R_{Ohm}$ has intermediate error bars. In general, the error bars increase from Li over Na and K to Rb due to higher reactivity and dendrite growth rate. However, also the low solubility of K and Rb salts in carbonates increases the uncertainty due to less good electrolyte conduction and charge transfer.

A second comparison in Figure 6 shows that also the $H$ and $j_{0,H}$ parameters cluster for both Li and Na at each temperature. At 20°C one can observe a splitting of the seven different electrolytes of Li because 1 M LiTFSI and 1 M LiBOB in PC have their $T_C$ between 20–25°C while the other five Li electrolytes have their $T_C$ between 0–20°C. However, at 50°C all Li-based electrolytes cluster again as all are well above $T_C$. Exception from this general clustering is detected for $PF_6^-$ in PC and Na electrolytes at higher temperature which will be discussed in Part II of this series.

The exchange current $j_{0,BV}$ is very similar for all seven Li and three different Na electrolytes as shown in Figure 6, and also the fitted ohmic resistance matches literature in Figure 4 for all found electrolyte conductivities in literature. Thus, one can be relatively certain that the usage of these three equations unifies our understanding of alkali-ion batteries. The parameters converge for EIS and GS over seven orders of magnitude for the SEI resistance. Also, charge transfer of the electron hoping from the metal electrode to the alkali-ion is very similar for the same alkali-metal with a weak dependence on salt and solvent, however, a significant dependence on concentration and temperature.

### Discussion

***Behavior near critical temperature.***—The serendipity of finding a critical temperature for all seven tested Li-based electrolytes and one Na electrolyte is very difficult if not impossible to observe directly from the overpotential profiles. The temperature profiles of −10, 30 and 70°C of Figure 1 show basically no obvious anomaly at 0°C and 5°C which are just below and above the critical temperature $T_C$ (SI Fig. S12).

Figure 7 shows the evolution of the parameters of Li/Na-$PF_6$ in EC:DMC near their individual $T_C$. A zoom to only 0–1.5 mAcm$^{-2}$ has been chosen to improve visibility. At temperatures above $T_C$, the overpotential from the SEI is negligible. Between 0–5°C, and 50–60°C a sudden change of the overall contribution can be observed for 1 M LiPF$_6$ and 0.5 M NaPF$_6$ in EC:DMC, respectively. A precise $T_C$ cannot be given, first, because the temperature intervals are too broad being usually 10°C. Second, and even more important, $T_C$ is different for the initial and steady-state overpotential evaluation by a few degree as shown in Figure 4. For the case with the best temperature resolution of 1 M LiPF$_6$ and 1 M LiClO$_4$ both in EC:DMC, $T_C$ of the steady state overpotential is in the range of 1 to 4°C lower than the one determined from the initial overpotential. Thus, either there is fitting uncertainty or ionic conduction in the inner SEI is worse for thick stable SEIs after a 10 min open-circuit relaxation than for continuous built-up of new thin SEI layers on the freshly deposited alkali metal during galvanostatic cycling. Both possibilities are reasonable.

The charge transfer reaction often compensates for the rapid changes of the SEI transport near $T_C$ which can be seen by jumps of $j_{0,BV}$ in most of the seven Li-based electrolytes. However, far below $T_C$, Butler-Volmer cannot compensate anymore for the rapidly increasing experimental overpotential. One needs to stress that uncertainty for the very small overpotential contribution in the case of Li could have originated easily from fluctuation or systematic errors. But the match of the SEI resistance from EIS with the one from GS for all four alkali metals over orders of magnitude, the higher $T_C$ of Na compared to Li and the overall clustering of parameters in general for different electrolytes could be an indicator that there might be common transport mechanisms for all alkali-ions.

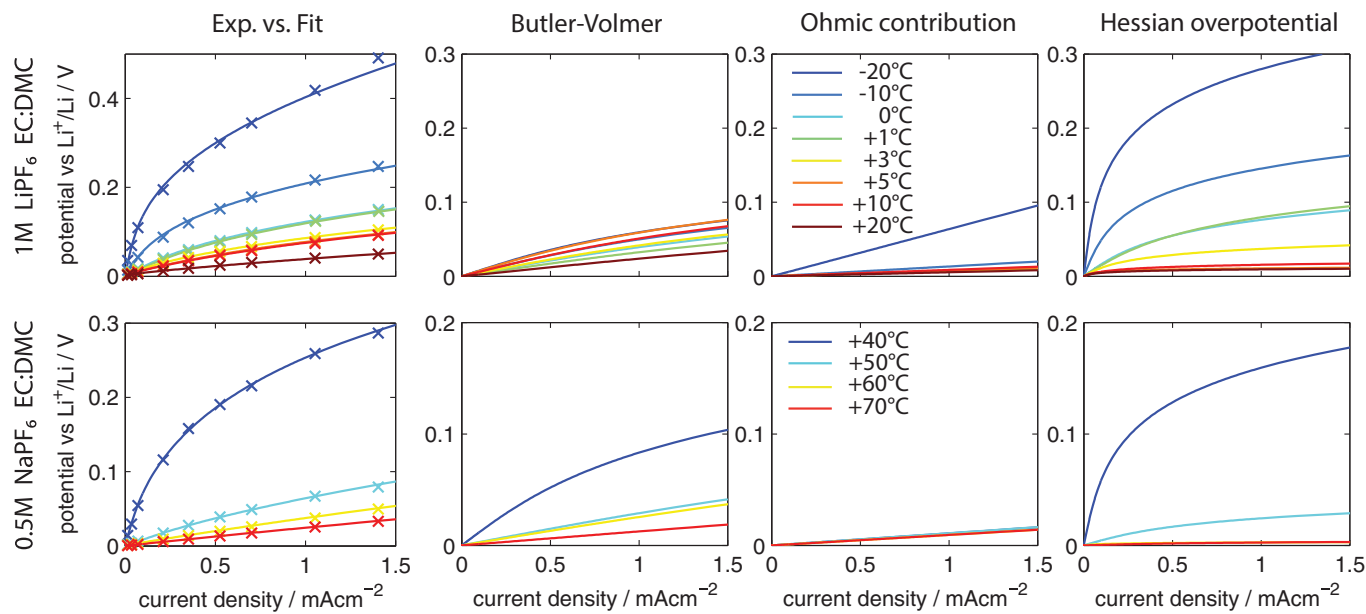

**Figure 7.** Overpotential increase and individual transport processes evolution near critical temperature: two different electrolyte systems of 1 M LiPF6 and 0.5 M NaPF6 in EC:DMC chosen where charge transfer is major overpotential source at temperatures above $T_C$ ($T>5°C$ for LiPF$_6$ and $T>50°C$ for NaPF$_6$) but the ionic conduction in the SEI becomes the main limitation far below $T_C$ ($T<0°C$ and $T<40°C$); while in all seven Li-based electrolytes the charge transfer reaction compensates with parameter jumps of $j_{0,BV}$ for the rapidly changing $H$ and $j_{0,H}$ parameter near $T_C$, $j_{0,BV}$ in the NaPF$_6$ EC:DMC case cannot account for such dramatic change of the SEI resistance anymore; however, uncertainty persists due to significant assumptions made to evaluate the data.

However, one has to keep in mind that dendrite growth, asymmetries of Butler-Volmer and Hess equation, and changes of the electrodes during cycling cannot be compensated in the symmetric cell setup tested here. Also, if one choses another fitting algorithm with equally weighted current densities (e.g 10 mV difference at 50 and 0.01 mA weighted equally), one fails to extract consistent parameters for Equation 3 (SI Fig. S14). This is due to the sensitivity of the equation to low and very low current densities only (SI Fig. S15). Thus, further research needs to be conducted to reveal the underlying transport mechanism.

*Distinct critical temperature.*—The author of this publication is very sceptic with regard to a precise phase transition with a defined critical temperature. The SEI is made out of various deposition products which react non-selectively once the alkali-metal is immersed in an electrolyte. While there seem to be common reduction products in the outer and inner SEI,[32] their crystallinity, grain boundaries, porosity distribution and polymer chain length should vary widely. Thus, it is very questionable if a distinct phase transition occurs uniformly over the entire area of the SEI.

A rather continuous transition with variations due to impurities, age, concentration differences, and applied current density is more likely to be present. But the two assumptions of just three transport processes being all in series and a perfect symmetry might be too restrict to allow the fitting process a smoother transition. Therefore, the hypothesis of two different transport mechanisms in the SEI in parallel (e.g. bulk and grain boundary ion conduction) and simply two different SEIs on the charged and discharged alkali metal electrode have been tested. And indeed, the addition of a second transport either in series (assumption $i_{BV} = i_{Ohm} = i_{H1} = i_{H2}$) or in parallel to the first Hess equation (assumption $i_{BV} = i_{Ohm} = i_{H1} + i_{H2}$) resulted in a good fit near the critical temperature especially for the parallel model (SI Fig. S16). However, the features of the overpotential to current density profiles (e.g. seen in Fig. 2) are simply too weak to be able to determine six fitting parameters independently. Thus, there remains uncertainty near the critical temperature.

But one aspect is very clear. The overpotential of the SEI changes to highly resistive values at low temperatures. This is validated by GS and EIS, independent of the fitting procedure (SI Note 5) and independent of any assumption made to evaluate its origin. While the muting factor $H$ can already merge the EIS and GS data over seven orders of magnitude, another approach might further simplify the present model. Therefore, other researchers are encouraged to investigate the SEI overpotentials and help to shed some light on this particular issue in battery research.

*Mechanisms of ionic conduction.*—Assuming that systematic errors do not influence the extracted parameters significantly and that a phase transition with a destinct critical temperatures does exist, one can analyze the underlying ionic transport mechanism. On the one hand, there seems to be a common transport mechanism through the SEI for Li, Na, K and Rb below $T_C$ which depends mainly on the conducted alkali-ion ($1/R_{Hess}$ in Figure 5). On the other hand, a sudden jump of the $H$ and $j_{0,H}$ parameter over 1.5 to two orders of magnitude for all seven Li electrolyte systems and one out of four Na systems shows another contribution. Three possibilities seem to be able to explain this anomaly.

First, a significant asymmetry of the deposition and dissolution reaction both in the Butler-Volmer and Hessian equation together with significant dendrite growth could increase the surface area for $j_{0,BV}$ and $j_{0,H}$. This could lead to two very different morphologies of the alkali electrodes in the symmetric cells. The basic assumption of symmetry would, thus, recognize two different processes artificially. Only a reference electrode or asymmetric cells with e.g. graphite or TiS$_2$ counter electrodes could be able to quantify the asymmetry.

Second, the jump of $H$ and $j_{0,H}$ could be associated with a first order phase transition of one process alone at a critical temperature $T_C$. Below $T_C$ only slow conduction would be possible while above $T_C$ the ions can be transported easily.

The third possibility is based on two different transport mechanisms through the inner dense SEI which are in parallel. One mechanism can conduct alkali ions very well but undergoes a phase transition during some kind of reordering of its underlying atomic configuration or a glass transition at $T_C$. The second process conducts ions at all temperatures from −40 to +80°C but with a much higher associated overpotential/energy for ion hopping. This could be, e.g., bulk ionic conduction in the grains and in parallel grain boundary conduction where $j_{0,H}$ would be much smaller for the better conducting grain boundary conduction which would be in line with the jump too lower $j_{0,H}$ values above $T_C$ in Figure 4. Thus, with the data presented here, this parallel conduction of alkali-ions in the bulk at all temperatures and along the grain boundaries above $T_C$ seems to be the most plausible one.

What is the error if we have two parallel processes but use just one process to fit the data to get Figure 4. Let us consider that at a certain temperature sufficiently far from $T_C$, mainly one process dominates due to its significantly smaller ionic resistance. Thus, if both resistances are in parallel and e.g. one process has 10x smaller ionic resistance it carries 91% of the current. If we have two orders of magnitude difference, as we observe here, the better process carries already 99% of the current. Thus, sufficiently far from $T_C$, the basic assumption $i_{BV} = i_{Ohm} = i_{Hess}$ would hold and the data analysis presented in this paper would have only 1% error.

The other extreme case considers operation directly at $T_C$ where both process have similar resistances. Thus, the total resistance of both processes would be just half of each individual one. However, assuming that a single process carries all current and not just half would result in an underestimation of the resistance by a factor of two. This could explain the underestimation of the Hessian resistance near $T_C$ in Figure 4.

Calculations based on two different SEI resistances a) in series and b) in parallel for the possibilities one and three are simulated in SI Note 6. The transport of ions in two different but parallel ion transport mechanisms fits the experimental data best near the critical temperature. Therefore, it is likely that two different transport mechanisms exist where the more efficiently one shuts down at a certain $T_C$ and only the basic transport mechanism is left. While $T_C$ could be determined for all Li electrolytes and one Na electrolyte, the other three Na-electrolytes and the K and Rb cells only show one transport mechanism. However, from the increase of $T_C$ from Li to Na, one would expect $T_C$ to be above the melting point of K and Rb-metal and therefore could not be tested in the current setup. The comparison of the $H$ parameter with the Young equation[14,15] yields non-physical SEI thicknesses in the range of Å for both of the two ionic conduction processes. Thus, further research is necessary to detect the origin of the underlying transport mechanisms and their associated SEI overpotentials.

## Conclusions

Investigation of symmetrical electrodes of Li, Na, K, and Rb metal confirms that the overpotential of the solid-electrolyte interphase is strongly non-linear over the temperature range from $-40°C$ up to $80°C$ for carbonate based electrolytes. The empirical equation proposed earlier,[1] can merge the non-linear SEI resistance from galvanostatic experiments in the range from a few mV for Li up to several V for Rb to the first semi-circle in the Nyquist plots from electrochemical impedance spectroscopy ranging from circa 26 $\Omega cm^2$ for Li up to 292 $M\Omega cm^2$ for Rb. The match of the ohmic resistance in the electrolyte with literature data of precise determined electrolyte conductivities and the overall similarity of the exchange current densities of Butler-Volmer for the same alkali-metal in various different electrolyte systems seems to confirm that the currently proposed equation for the ionic transport within the inner SEI might unify the understanding of alkali-metal electrodes.

Also indicators exist for a possible phase transition in the SEI for all seven Li-based electrolytes and one out of four Na electrolytes. The associated critical temperatures are between 0–25°C for the seven Li electrolytes and between 50–60°C for the Na one. While no direct proof of a phase transition can be given here, the measured changes in the associated overpotential are significant. While the current publication confirms that the SEI overpotentials for Li electrodes are negligible at room temperature;[1] these overpotentials become dominant at very low temperature confirmed both by galvanostatic and impedance spectroscopy. The current study also confirms that K and Rb would possess low ionic conduction within the SEI which results in significant overpotentials. However, also Na has low conduction at room temperature and a much higher critical temperature above 50°C, which places certain disadvantages compared to Li-based electrodes.

Overall, further investigation of the conduction mechanisms within the SEI and the possibility of a phase transition with a critical temperature for e.g. the SEI on graphite, would be of outmost importance for battery management systems. Additionally, assuming a zero SEI resistance in battery models would results in very little error during parameter estimation if one fits data above $T_C$. However, any state-of-health and state-of-charge estimator will determine incorrect parameters below $T_C$ and might cause significant safety issues during recharging of Li-ion based batteries at low temperatures which would be a critical issue for the electrification of the transportation sector.

## Acknowledgments

I would like to thank the reviewers for helpful comments on the manuscript. Additionally, I thank V. Wood, the Swiss National Science Foundation (grant 20PC21_15566/1) and Battronics (project NonLin-SEI) for the financial support without this study would not have been possible to conduct.

## ORCID

Michael Hess  https://orcid.org/0000-0001-9921-4735

# Supplementary Information to: Temperature-dependence of the solid-electrolyte-interphase overpotential: Part I – two parallel mechanisms, one phase transition

Michael Hess[1,2,*]

[1]ETH Zurich, Laboratory of Nanoelectronics, Department of Electrical Engineering and Information Technology, 8092 Zurich, Switzerland

[2]Battronics AG, β-research division, 8037 Zurich, Switzerland

**ABSTRACT:** Here, specific information and individual data for each of the different electrolyte systems at each temperature is given for the case of galvanostatic and electrochemical impedance spectroscopy.

## 1. Supplementary Note 1: Experimental

The salt was added based on the calculated density of the pure solvent mixture meaning that the density change by the salt addition was neglected due to insufficient information about the density of 1M solutions of the respective electrolytes. This leads to a systematic error of having only 0.961-0.966M electrolytes for the 1M $LiPF_6$ in EC:X (X=DMC,EMC,DEC) equivalent electrolytes where the densities of the solvent mixtures and electrolyte mixtures are well reported [1, 2]. The other electrolytes are expected to have a similar error of 3-4%.

Characterization of Separator: Commercial Whatman glass microfiber filters (GE Healthcare, GF/D 1823-257) of thickness 679±21 µm at zero pressure weighting 41.21±0.09 mg for 20 mm diameter punched samples were exclusively used (average of five test samples). In the uncompressed state they possess circa 92% porosity. After compression to circa 430±5 µm, they still possess 88% of porosity. The pressure in the cell is p=50 $Ncm^{-2}$ in the coin type cells. Additionally, the high transference number $t_{Li}$ =0.56 [3] is of advantage due to $PF_6^-$ trapping on the $SiO_2$ surface groups compared to inert PE or PP separators with $t_{Li}$ = 0.4 [4]. The glass fiber separator was heated at 400°C inside the glovebox to remove adsorbed water. The calculated resistance based on the thickness, porosity, conductivity for 1M $LiPF_6$ EC:DMC 1:1 [5] and an assumed tortuosity of 1.2, yields 3.93±0.04 Ω for EIS test.

Cycling protocols: For galvanostatic tests, all symmetrical cells were first cycled at very small current densities of 2.8, 7, and 14 $µAcm^{-2}$ to allow the formation of a homogeneous SEI. After these three cycles, two high current density "discharges" of 28 $mAcm^{-2}$ were used for which the overpotential changes significantly during the respective discharge which cannot solely be attributed to a surface area increase by dendrites but seems to change the ionic conduction or thickness of the SEI significantly. 28$mAcm^{-2}$ corresponds to the calculated limiting current density of the glass fiber separator. After this procedure several rates were tested starting from the highest current density of 56 $mAcm^{-2}$ to the lowest of 14 $µAcm^{-2}$. The respective "charge" was always performed at low current density of 14 $µAcm^{-2}$. To test the influence of the cycling protocol on the galvanostatic overpotentials, also protocols from low current density to high, and symmetrical current densities of "charge" and "discharge" current being the same have been tested previously and show a difference especially at the initial overpotential [6].

For electrochemical impedance spectroscopy, also the first five initiation cycles were applied being the three low current density SEI formation cycles and two high current density SEI activation cycles. After these cycles, one cycle at low current density of 14 $µAcm^{-2}$ was performed to "smooth" the surfaces from any dendrites to decrease the possibility of surface changes during EIS excitation. A 10h hold at open-circuit-potential was done before excitation with EIS at [2:2:20],[25:5:80],[90:10:200] mV in ascending order. This was done to test linearity of the EIS around 0V. All electrochemical tests were done with Biologic VMP3 and MPG2 cyclers between -40 to +80°C.

## 2. Supplementary Note 2: Individual electrolytes

Suppl. Fig. S1-S4 show the experimental GS data, best fit and individual contributions of charge transfer, ohmic resistance and SEI conduction modeled by Butler-Volmer, Ohms law and Hessian equation (eq. 1-3 of MM) for Li, Na, K and Rb, respectively. Suppl. Fig. S5-S6 show corresponding EIS data.

Suppl. Fig. S7-S10 show the evolution of the four parameters $j_{0,BV}$, $R_{Ohm}$, $H$ and $j_{0,H}$ of eq. 1-3 of MM over temperature in Arrhenius plots for each individual out of the sixteen different tested electrolyte systems. Note, that experimental data from literature has been added for the plots of $R_{Ohm}$ wherever possible to allow comparison of the fitted value to precise literature data. The exchange current density $j_{0,BV}$ varies widely in literature due to many wrong measurements and assumptions based on Tafel regimes etc. and is therefore not compared. Also the derived parameter $R_{Hess}$ has been added to allow comparison between the $H$ and $j_{0,H}$ parameter extracted from GS measurements and the resistance of the first semi-circle during EIS for all electrolyte systems. See eq. 6 of MM for details.

### 2.1 Fit to initial overpotentials $E_{init}$ of GS cycling

See Suppl. Fig. S1-S4 for fit to initial overpotential. The steady state overpotential was evaluated in the same way.

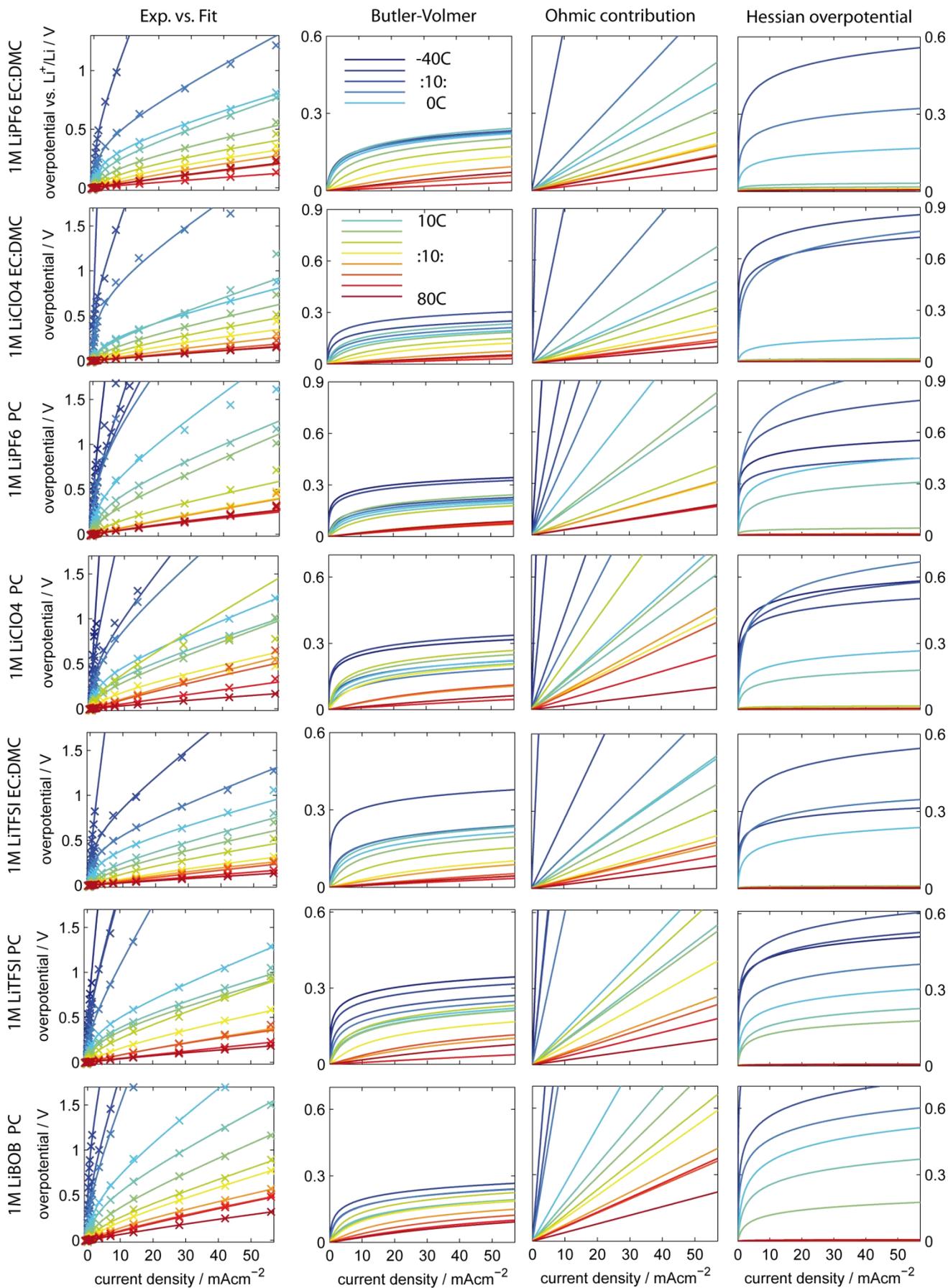

**Supplementary Figure S1: Evolution of overpotentials of all measured Li electrolytes:** experiments versus fit from -40 to +80°C for 1M salts of LiPF$_6$, LiClO$_4$, LiTFSI, LiBOB in either EC:DMC 1:1wt or pure PC with the respective contributions from the equations of Butler-Volmer, Ohm's law and Hess as a rainbow plot from dark blue for -40°C to dark red for +80°C.

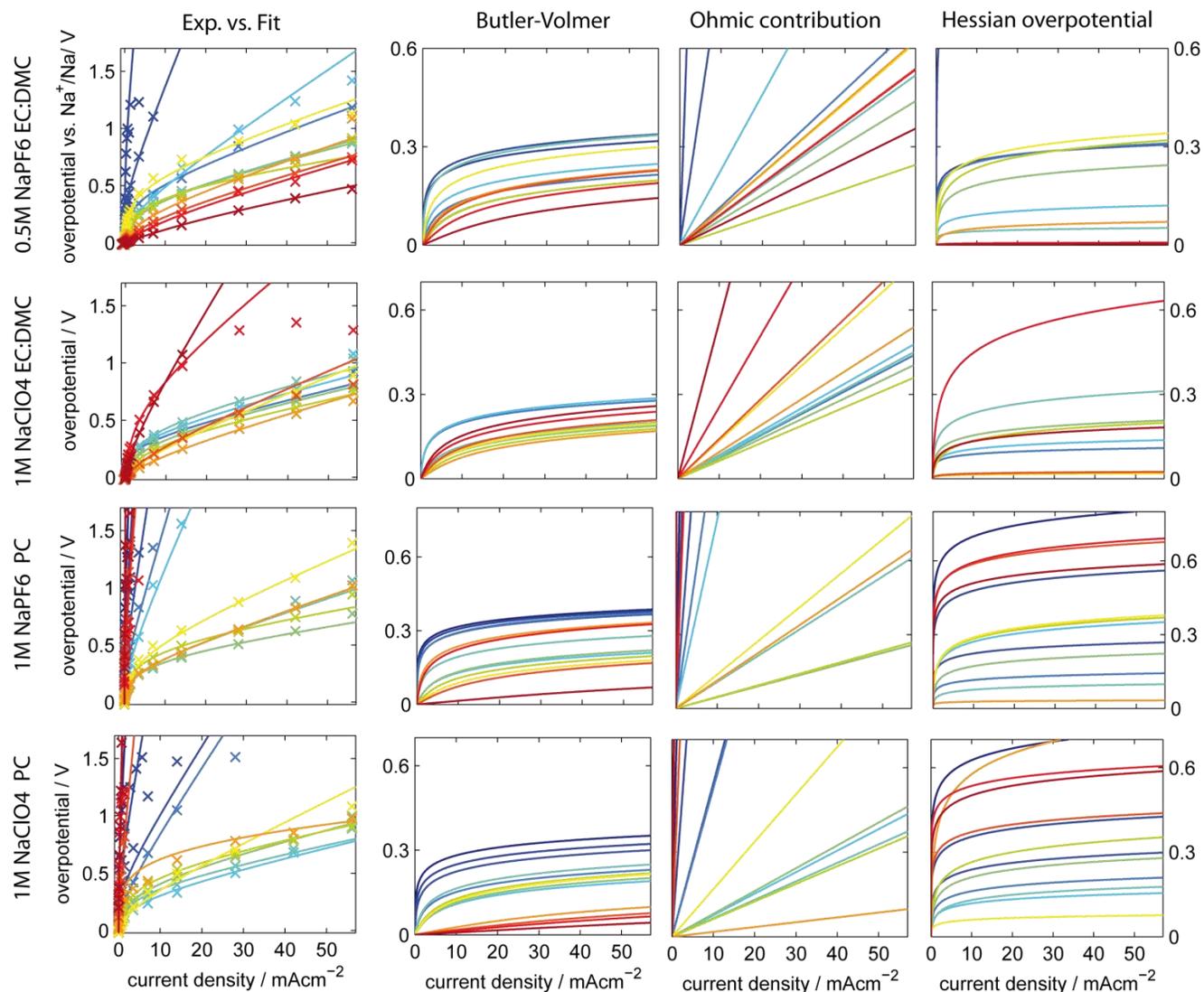

**Supplementary Figure S2: Evolution of overpotentials of all measured Na electrolytes:** experiments versus fit from -40 to +80°C for 0.5 or 1M salts of $NaPF_6$ and $NaClO_4$ in either EC:DMC 1:1wt or pure PC with the respective contributions from the equations of Butler-Volmer, Ohm's law and Hess as a rainbow plot from dark blue for -40°C to dark red for +80°C.

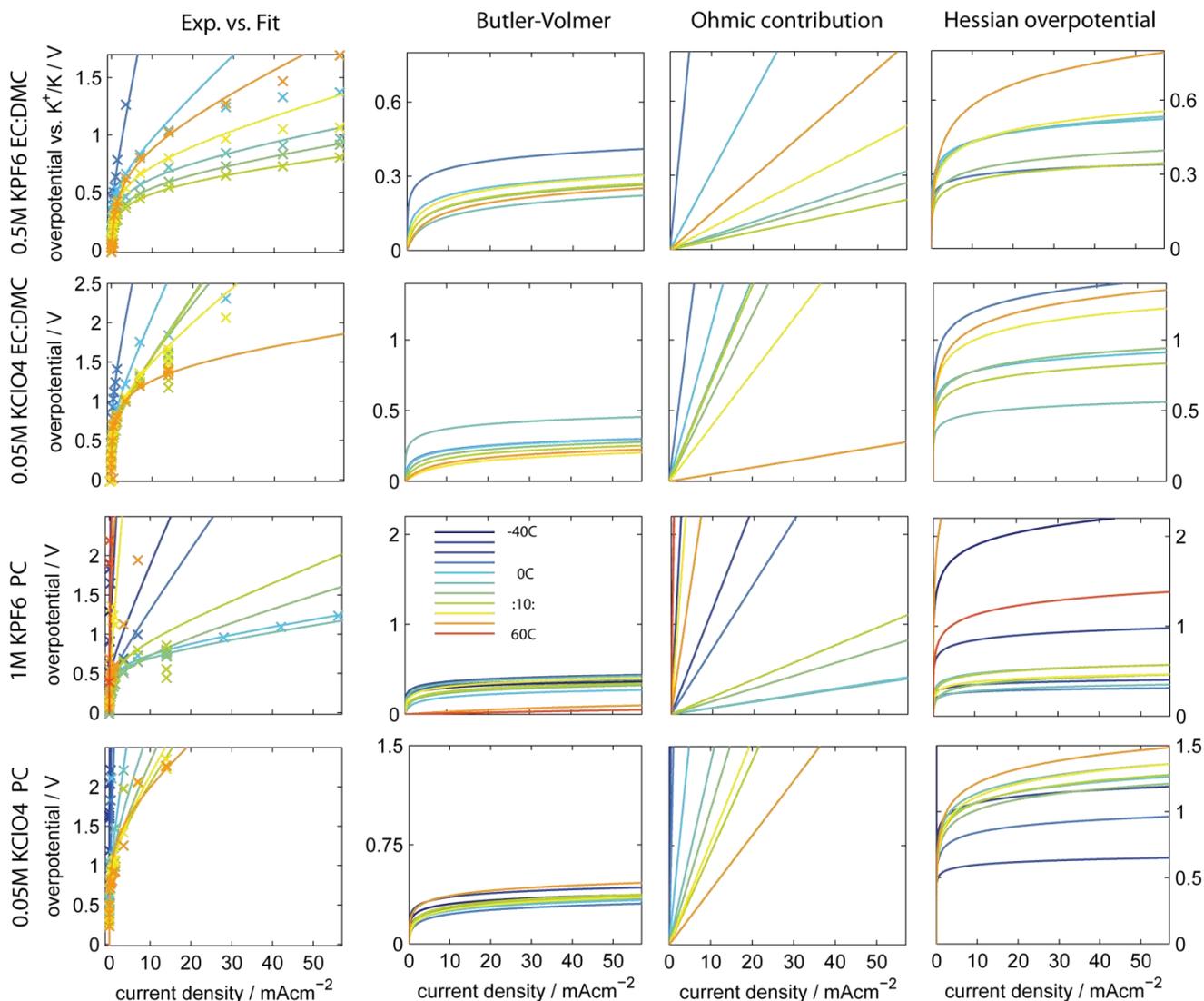

**Supplementary Figure S3: Evolution of overpotentials of all measured K electrolytes:** experiments versus fit from -40 to +80°C for 0.5 or 1M salts of $KPF_6$ and 0.05M $KClO_4$ in either EC:DMC 1:1wt or pure PC with the respective contributions from the equations of Butler-Volmer, Ohm's law and Hess as a rainbow plot from dark blue for -40°C to orange for +50°C or even red for +60°C, only 1M $KPF_6$ PC was possible to measure at 60°C where the other symmetric K-K cells had already short circuits due to the melting point of K at 63.5°C.

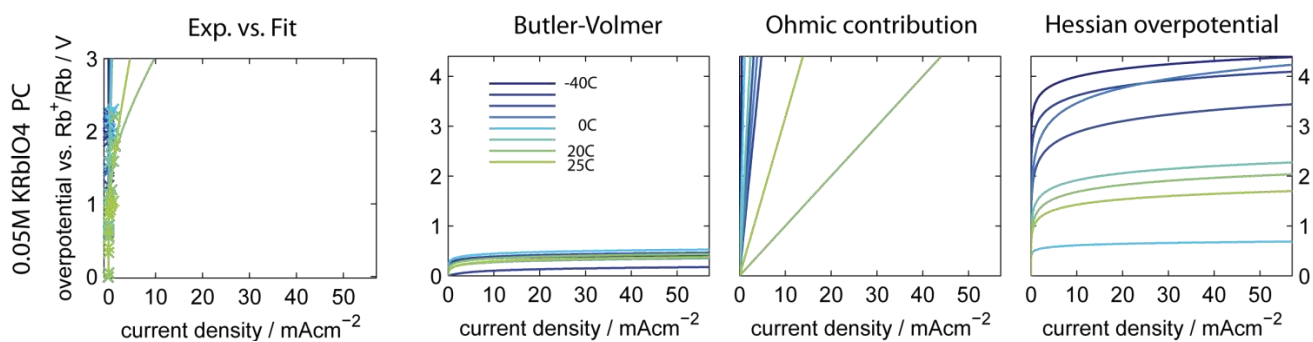

**Supplementary Figure S4: Evolution of overpotentials of 0.05M $RbClO_4$ in PC:** experiments versus fit from -40 to +25°C for 0.05M $RbClO_4$ in pure PC with the respective contributions from the equations of Butler-Volmer, Ohm's law and Hess as a rainbow plot from dark blue for -40°C to green for +25°C, test at 30°C of symmetric Rb-Rb cells had already short circuits due to the melting point of Rb at 39.3°C, reproducibility of Rb cells very bad due to low solubility of $RbClO_4$ in carbonates and limited applicable current density due to low electrolyte conductivity of 0.05M salt and the overwhelmingly high SEI resistance also seen in EIS experiments.

## 2.2 Fit to EIS data at 10 mV excitation

Suppl. Fig. S5-S6 show the EIS data plotted once in standard Nyquist plots with lin-lin scales and once in log-log plots for better visibility due to several orders of magnitude differences.

Note that the other data at excitation from 2:100 mV amplitude was not used for the evaluation here but linearity was observed up to relative high excitation but temperature dependent.

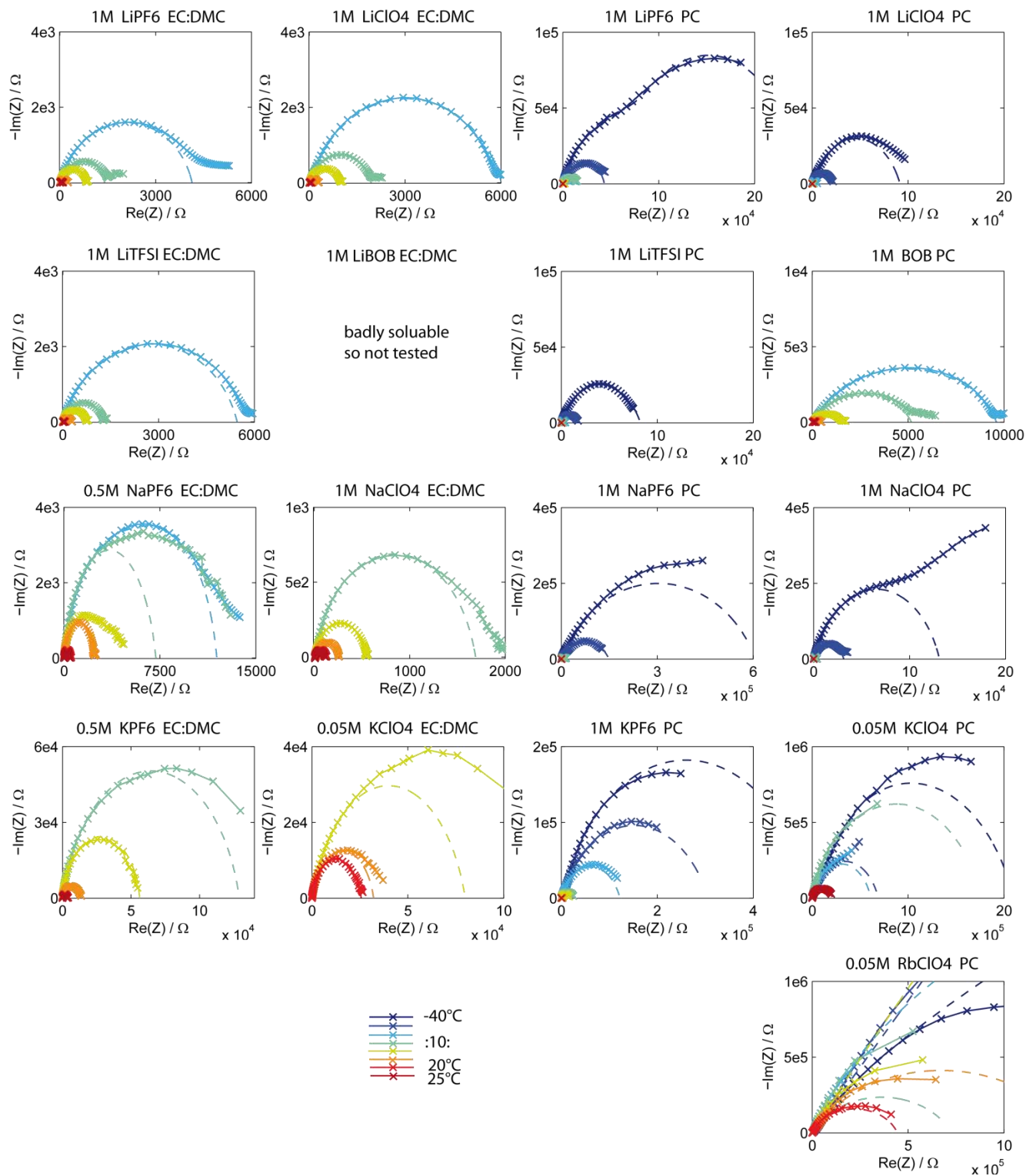

**Supplementary Figure S5: Experimental data and best fit of electrochemical impedance spectroscopy data in Nyquist plots:** all sixteen different electrolyte systems tested from minimum -40°C for PC samples only up to maximum 25°C; dashed line represents best fit to the first semi-circle based on an equivalent circuit of $R_1+R_2||Q_2$ or $R_1+Q_2||(R_2+R_3||Q_3)$; usually most simple circuit used; however; sometimes a small pre-circle at high frequency needs to be included to fit the standard very big semi-circle; no data on the low frequency part was evaluated as unphysical high surface areas and low exponents near 0.5 for the phase element indicate severe issues with either the data or the assumptions of using linear RQ elements as shown previously [6].

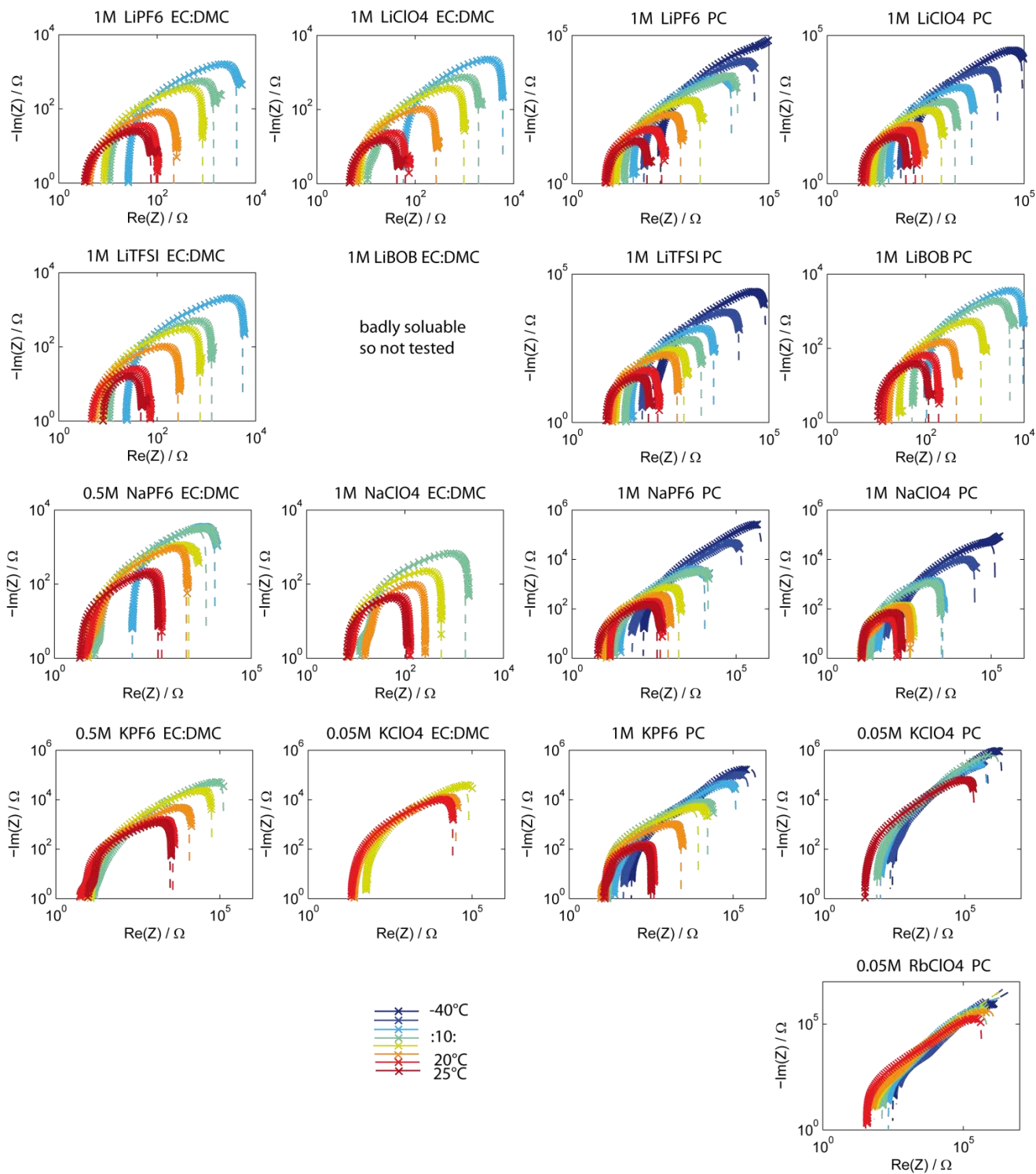

**Supplementary Figure S6: Experimental data and best fit of electrochemical impedance spectroscopy data in loglog plots:** the exact same data as in Suppl. Fig. S5 but plotted in log scale for better visibility of the evolution of the Ohmic resistance of the electrolyte at the intercept at high frequency and the evolution of the SEI resistance at medium to small frequencies; note the changes of electrolyte resistance are usually within one order of magnitude while the SEI resistance changes between two up to four orders of magnitude depending on electrolyte.

## 2.3 Arrhenius plots of estimated parameters

The following data is the same data as displayed in Figure 4 of MM, however, plotted individually to allow better comparison for each electrolyte system. Parameter estimation near $T_C$ resulted in large fluctuations of $H$ and $j_{0,H}$ of up to four tested samples at the same temperature. Additionally, $j_{0,BV}$ often compensated partially the significant changes of $H$ and $j_{0,H}$ while the Ohmic resistance was rather stable near $T_C$ and not influences much. This compensation is much more significant for EC:DMC based electrolytes and significantly less pronounced in PC electrolytes.

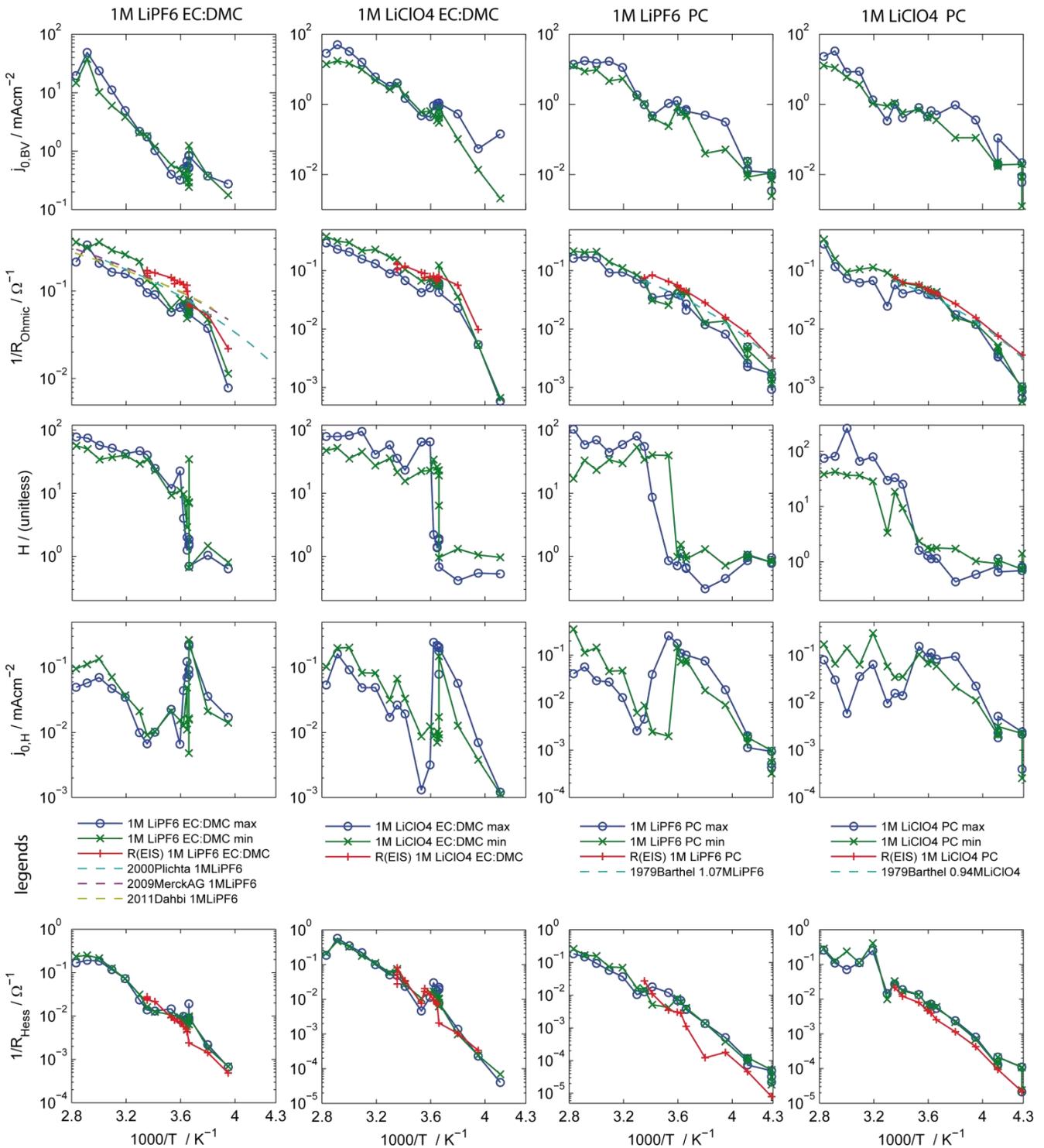

**Supplementary Figure S7: Individual parameter sets of best fit of symmetrical Li cells:** for the three processes Butler-Volmer, Ohm and Hess with **columns:** for 1M LiPF$_6$ and 1M LiClO$_4$ in either EC:DMC 1:1 or PC **rows:** 1) exchange current density $j_{0,BV}$, 2) ohmic resistance of GS and EIS experiments, 3) muting factor $H$, 4) exchange current density $j_{0,H}$ which are all four independent variables while row 6) plots the derived SEI resistance $R_{Hess}$ combining $H$ and $j_{0,H}$ in eq. 6 of MM showing the match with the SEI resistance from EIS experiments; electrolyte conductivity data of 1M LiPF$_6$ in EC:DMC [2, 5, 7], 1.07M LiPF$_6$ in PC [8] and 0.94M LiClO$_4$ in PC [8] has been added in row two calculated for the properties of the GF separator to show the match.

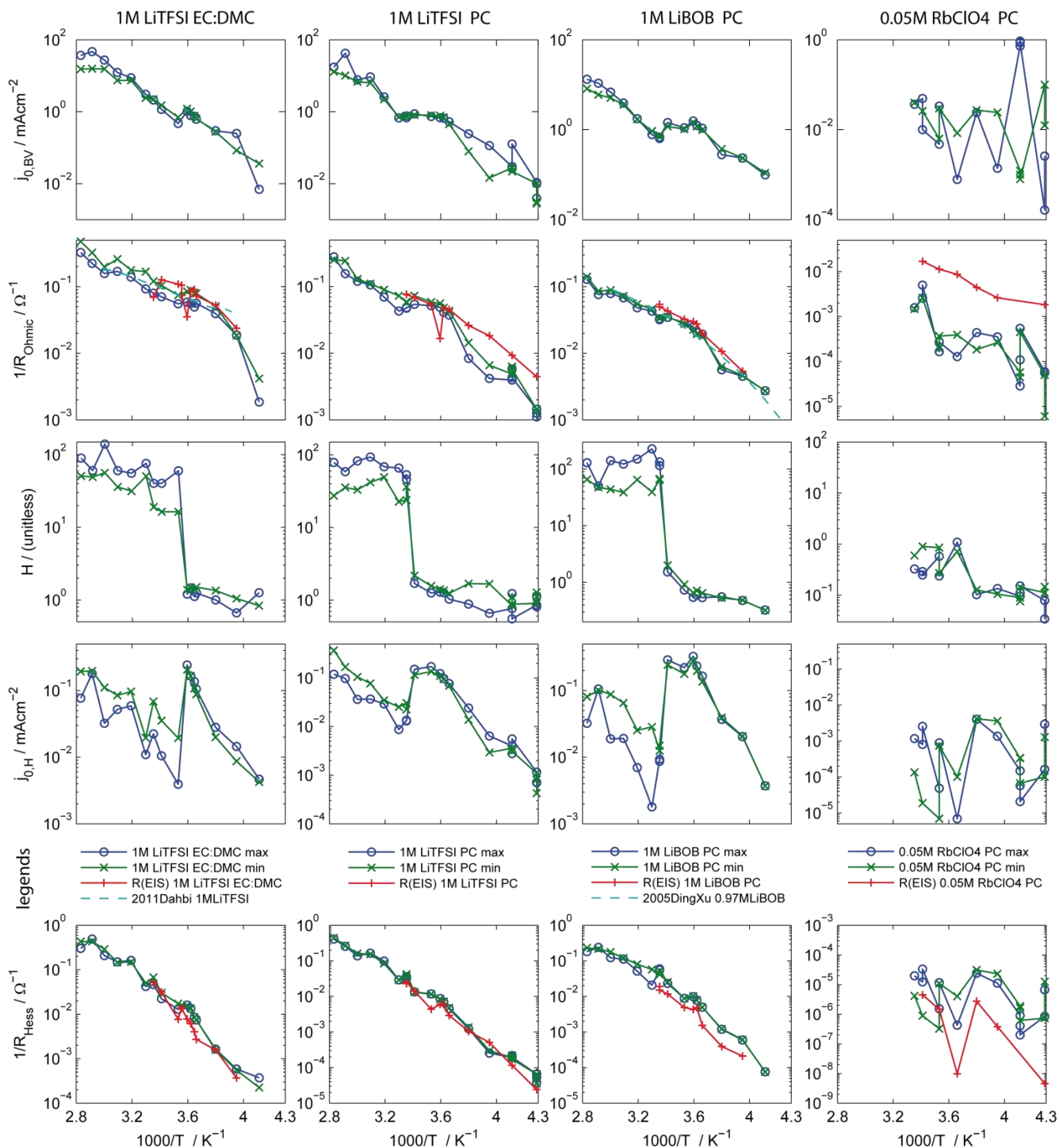

**Supplementary Figure S8: Individual parameter sets of best fit of symmetrical Li and Rb cells:** for the three processes Butler-Volmer, Ohm and Hess with **columns:** for 1M LiTFSI and 1M LiBOB in either EC:DMC 1:1 or PC and 0.05M RbClO4 in PC, **rows:** 1) exchange current density $j_{0,BV}$, 2) ohmic resistance of GS and EIS experiments, 3) muting factor $H$, 4) exchange current density $j_{0,H}$ which are all four independent variables while row 6) plots the derived SEI resistance $R_{Hess}$ combining $H$ and $j_{0,H}$ as shown in eq. 6 of MM showing the match with the SEI resistance from EIS experiments; electrolyte conductivity data of 1M LiTFSI in EC:DMC [7] and 1M LiBOB in PC [9] has been added in row two to show the match.

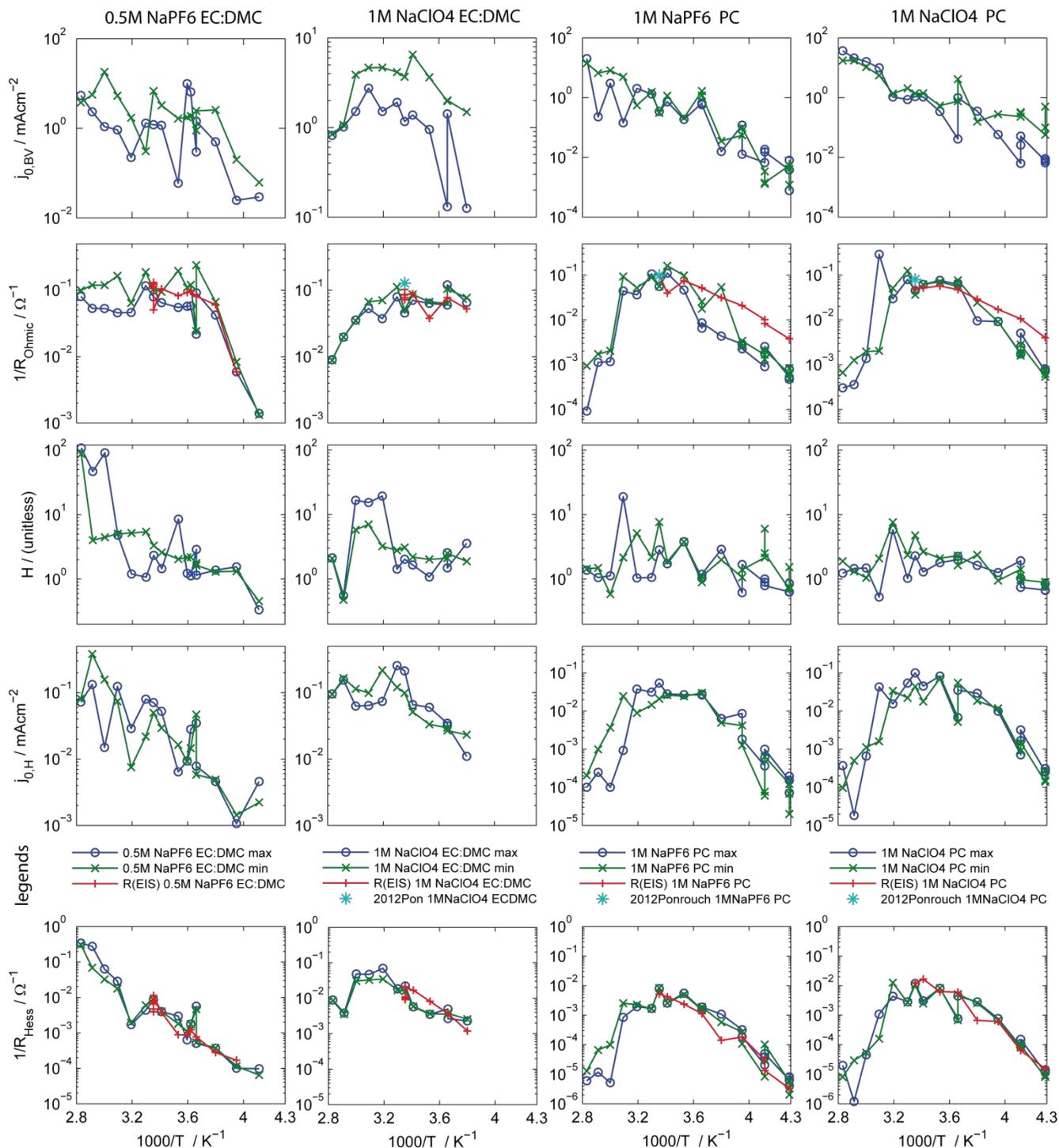

**Supplementary Figure S9: Individual parameter sets of best fit of symmetrical Na cells:** for the three processes Butler-Volmer, Ohm and Hess with **columns:** for 0.5/1M $NaPF_6$ or 1M $NaClO_4$ in either EC:DMC 1:1 or PC, **rows:** 1) exchange current density $j_{0,BV}$, 2) ohmic resistance of GS and EIS experiments, 3) muting factor $H$, 4) exchange current density $j_{0,H}$ which are all four independent variables while row 6) plots the derived SEI resistance $R_{Hess}$ combining $H$ and $j_{0,H}$ as shown in eq. 6 of MM showing the match with the SEI resistance from EIS experiments; electrolyte conductivity data for 1M $NaClO_4$ in EC:DMC and PC and 1M $NaPF_6$ in PC at 25°C only [10] has been added in row two to show the match.

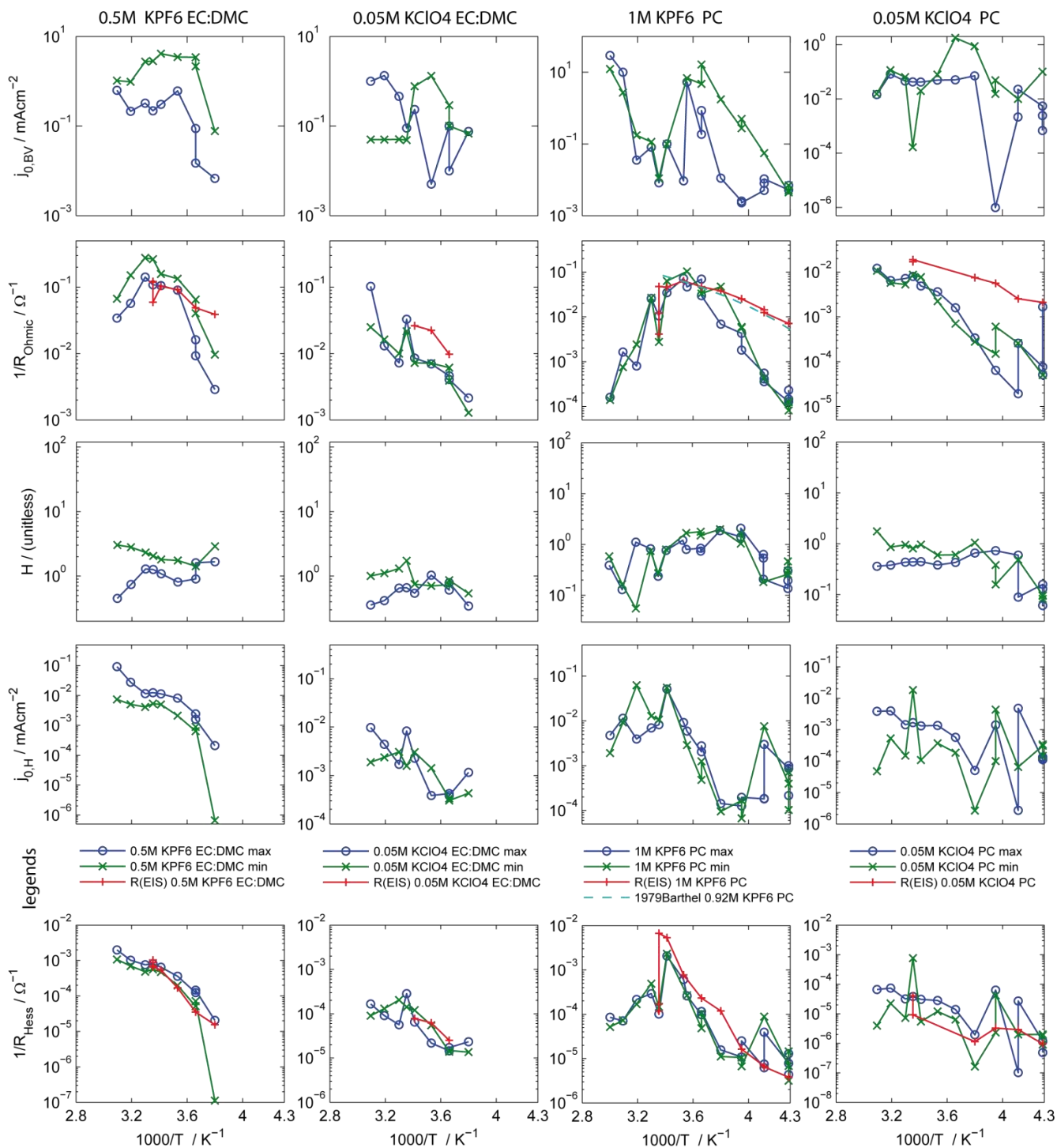

**Supplementary Figure S10: Individual parameter sets of best fit of symmetrical K cells:** for the three processes Butler-Volmer, Ohm and Hess with **columns:** for 0.5/1M $KPF_6$ or 0.05M $KClO_4$ in either EC:DMC 1:1 or PC, **rows:** 1) exchange current density $j_{0,BV}$, 2) ohmic resistance of GS and EIS experiments, 3) muting factor $H$, 4) exchange current density $j_{0,H}$ which are all four independent variables while row 6) plots the derived SEI resistance $R_{Hess}$ combining $H$ and $j_{0,H}$ as shown in eq. 6 of MM showing the match with the SEI resistance from EIS experiments; electrolyte conductivity data of 0.92M $KPF_6$ in PC [8] has been added in row two to show the match.

## 3. Suppl. Note 3: Error bars of parameter estimates

Evolution of the parameter estimates in Fig. 6 of MM shows the best fit parameter. The confidence interval calculated by Matlab's Levenberg-Marquart least square algorithm "fit()", are shown in Suppl. Fig. S11 for -10, 20 and 50°C.

If the error bars of $j_{0,BV}$ or $j_{0,H}$ do not end at the smallest value in the x-scale, the lower value is usually negative which would be infinite in logarithmic scale chosen here. Often the parameter $j_{0,BV}$ is insensitive and shows very large error bars over several orders of magnitude.

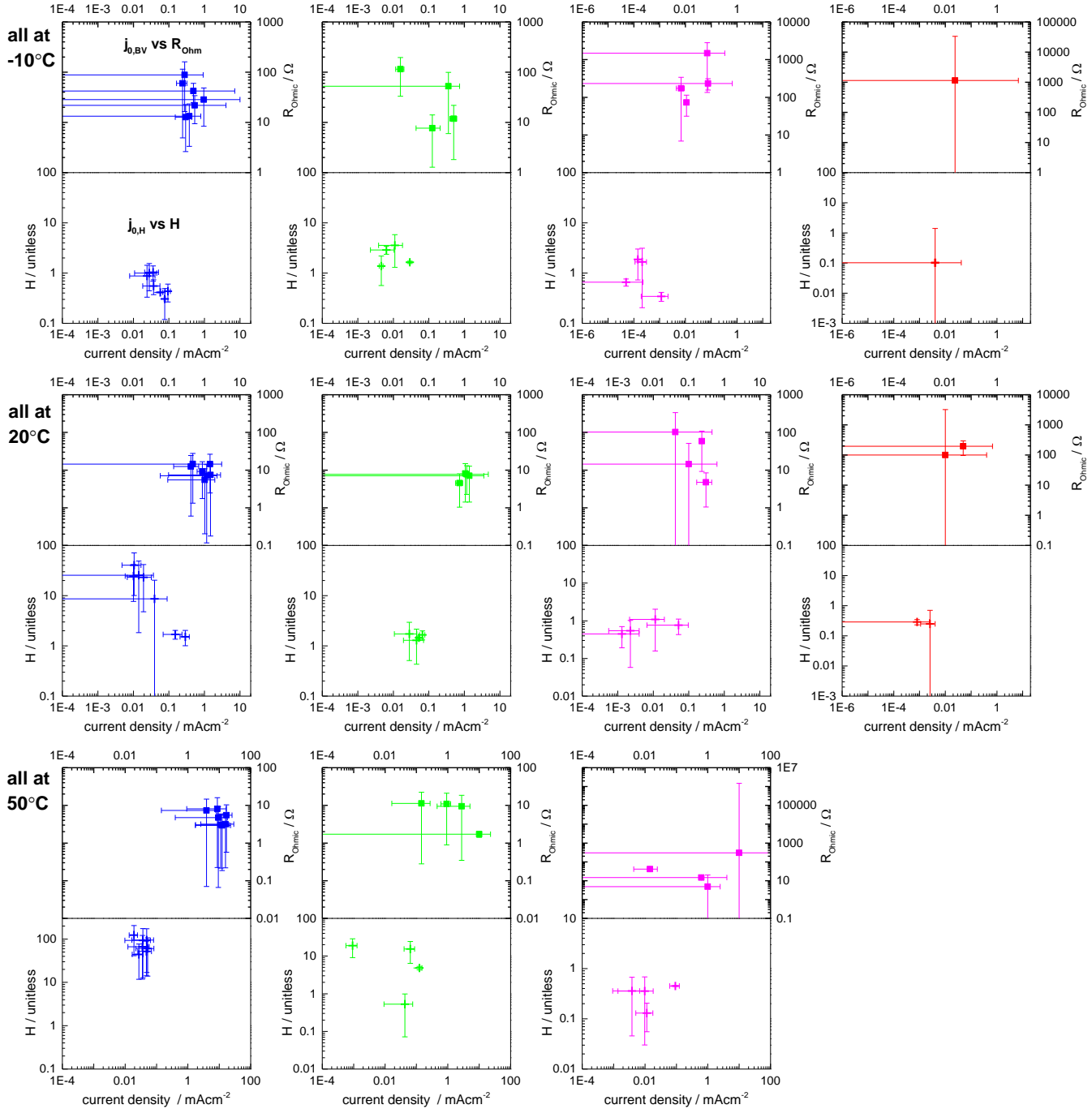

**Supplementary Figure S11: Error bars of confidence interval of Matlabs' Levenberg-Marquart fit for parameter evolution plots in Figure 6 in MM:** at -10, 20 and 50°C for all seven Li electrolytes (blue), the four Na electrolytes (green), four K electrolytes (magenta) and the single Rb electrolyte (red); parameters $j_{0,H}$ vs $H$ plotted in lower part and $j_{0,BV}$ vs $R_{Ohm}$ in upper part of split plot of Figure 6 in MM; error bar of $j_{0,H}$, $j_{0,BV}$, $H$ and $R_{Ohm}$ sometimes even negative resulting in -∞ for log scale so hit the left or lower bound of the axes in the plot, -10 and 50°C chosen as all EC:DMC based electrolytes still liquid while at 60°C three out of four K electrolytes died due to K- melting point at 63.5°C; **conclusion)** error bar of $j_{0,BV}$ usually the highest due to the main sensitivity at high current density where dendrites and electrolyte depletion effects contribute to systematic errors, $R_{Ohm}$ shows also relative high error bars despite the relative precise determination of the fitting value with low fluctuations between different nominally same samples, the muting parameter $H$ has the third highest error bars while $j_{0,H}$ is the most precise value to be determined despite its very low absolute value which would usually make it very prone to fluctuations; however, $j_{0,H}$ and $H$ seem to be the most sensitive values to extract during galvanostatic experiments due their main features at very low current density where dendrite growth, electrolyte depletion and asymmetry effects of Butler-Volmer are negligible.

## 4. Suppl. Note 4: Behavior near critical temperature

The serendipity of finding a critical temperature for all seven tested Li-based electrolytes and one Na electrolyte is very difficult if not impossible to observe directly from the overpotential profiles as shown in Suppl. Fig. S12 for 1M $LiPF_6$ EC:DMC 1:1wt.

The temperature profiles of -10, 30 and 70°C are the same as in Fig. 1 of MM. The profiles at 0°C and 5°C are just below and above the critical temperature $T_C$. It is basically impossible to detect any differences neither in the offset at the lowest current density of 0.7 mAcm$^{-2}$ (orange line) nor at any of the medium and high current densities.

However, if one fits the experimental overpotentials at both the initiation and in the near steady-state regime, one can observe a jump of the $H$ and $j_{0,H}$ parameters at certain $T_C$'s depending on alkali-metal, salt and solvent as shown in Suppl. Fig. S13. It needs to be stressed that the current density is zoomed to only 0-1.5 mAcm$^{-2}$ where the Hessian overpotential is most sensitive with steepness of its voltage rise based on $j_{0,H}$ and its saturation potential based on $H$. Both $j_{0,BV}$ and $R_{Ohm}$ are more sensitive and show more overall contribution at medium and high current densities up to the measured 56 mAcm$^{-2}$.

At the outmost right column of Suppl. Fig. S13, the Hessian overpotential is plotted alone mimicking the ionic conduction within the inner SEI. Between 3-5°C, 10-20°C, 20-25°C and 50-60°C a sudden change of the overall contribution can be observed for 1M $LiPF_6$ EC:DMC and PC, 1M LiBOB PC and 0.5M $NaPF_6$ EC:DMC, respectively. A precise $T_C$ cannot be given, first, because the temperature intervals are too broad being usually 10°C. Second, and even more important, $T_C$ is different for the initial and steady-state overpotential evaluation as shown in Fig. 4 of MM. For the case with the best temperature resolution of 1M $LiPF_6$ and 1M $LiClO_4$ both in EC:DMC, $T_C$ of the steady state overpotential is 1-4°C lower than the one determined from the initial overpotential. Thus, either there is fitting uncertainty or ionic conduction in the inner SEI is different for thick stable SEIs after a 10 min open-circuit relaxation and for continuous built-up of new thin SEI layers on the freshly deposited SEI during galvanostatic cycling. Both possibilities are likely.

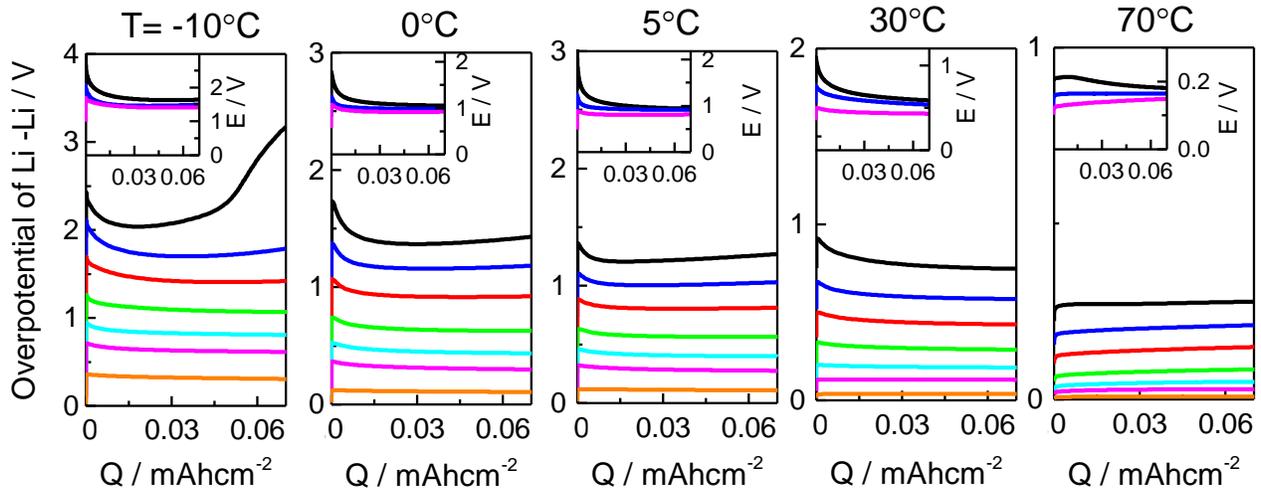

**Supplementary Figure S12: Similar plot as Figure 1 of MM of 1M $LiPF_6$ EC:DMC at different temperature:** include the overpotential evolution with current density just below critical temperature $T_C$ at 0°C and just above at 5°C, **in general)** no significant jump of the smallest current density of 0.7 mAcm$^{-2}$ in orange can be observed as would be expected if the SEI transport changes over one to two orders of magnitude.

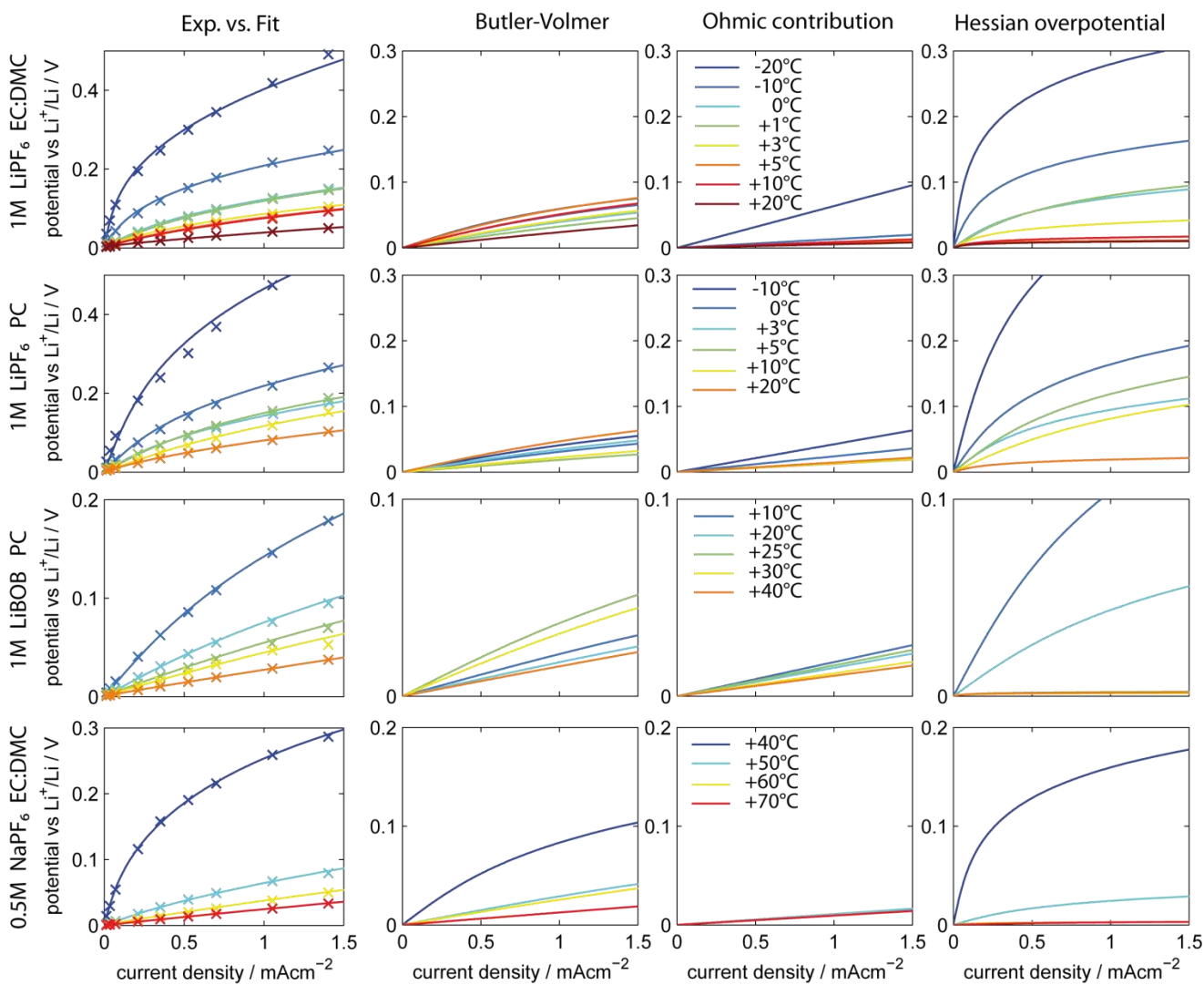

**Supplementary Figure S13: Parameter evolution near the critical temperature $T_C$ for four different systems:** including 1M LiPF$_6$ in EC:DMC or PC, 1M LiBOB in PC and 0.5M NaPF$_6$ in EC:DMC; experiment shown as markers x with best fit and individual contributions from charge transfer, Ohms law and SEI overpotential for different temperatures around each specific $T_C$ indicated in the legend; **conclusion)** while for 1M LiPF$_6$ in EC:DMC the Butler-Volmer overpotential improves again at $T<T_C$ (also seen as a jump in Fig. 4 of MM) which is simply due to the compensation effect during fitting and no real physical improvement, for 1M LiPF$_6$ and 1M LiBOB in PC this effect is already smaller as the SEI overpotential becomes more limiting and both Butler-Volmer and the Ohmic resistance can hardly compensate for the changes; in contrast for 0.5M NaPF$_6$ no compensation is anymore possible as the SEI overpotential is dominant for 40°C and 50°C but negligible from 60°C on; so for Li-metal, $T_C$ is difficult to determine exactly because the SEI overpotential is a minor contribution and small errors and fluctuations have a significant influence (high sensitivity to small fluctuations) while for Na-metal, the SEI overpotential is a major contribution to the overall overpotential and can be determined relatively accurately (low sensitivity to errors).

## 5. Suppl. Note 5: Other fitting algorithms

To illustrate the challenges with the fitting algorithm to the GS response, the standard least-square method is applied with equal weights of high and low current density overpotentials. Suppl. Fig. S14 shows the parameter evolution for the standard LSQ method. This means that a 10 mV deviation at 56 mAcm$^{-2}$ would be as important as a 10 mV difference at 0.018 mAcm$^{-2}$. However, this is electrochemically not justified as these 10 mV would correspond to an error of 50-200% at low but only 1.6-4% error at high current density.

However, here the standard LSQ method is shown in comparison to the applied method with high weights at low rates and very small weights at high current density used throughout this and the previous publication [6]. If one applies all data from charge and discharge at all rates, the method is strongly fluctuation as seen in column two of SI Fig. S14 and in green in column three and four. If one applies the method to the trusted region of just discharge measurements from low up to a certain high current density depending on temperature (14 mAcm$^{-2}$ below 0°C, 28 below 40-50°C, 42 mAcm$^{-2}$ up to 80°C) the parameters of $j_{0,BV}$ and $R_{Ohm}$ show similar values as in Fig. 3 of MM but $H$ and $j_{0,H}$ have intense fluctuations due to their sensitivity at low rates which are often not fit by the standard LSQ method.

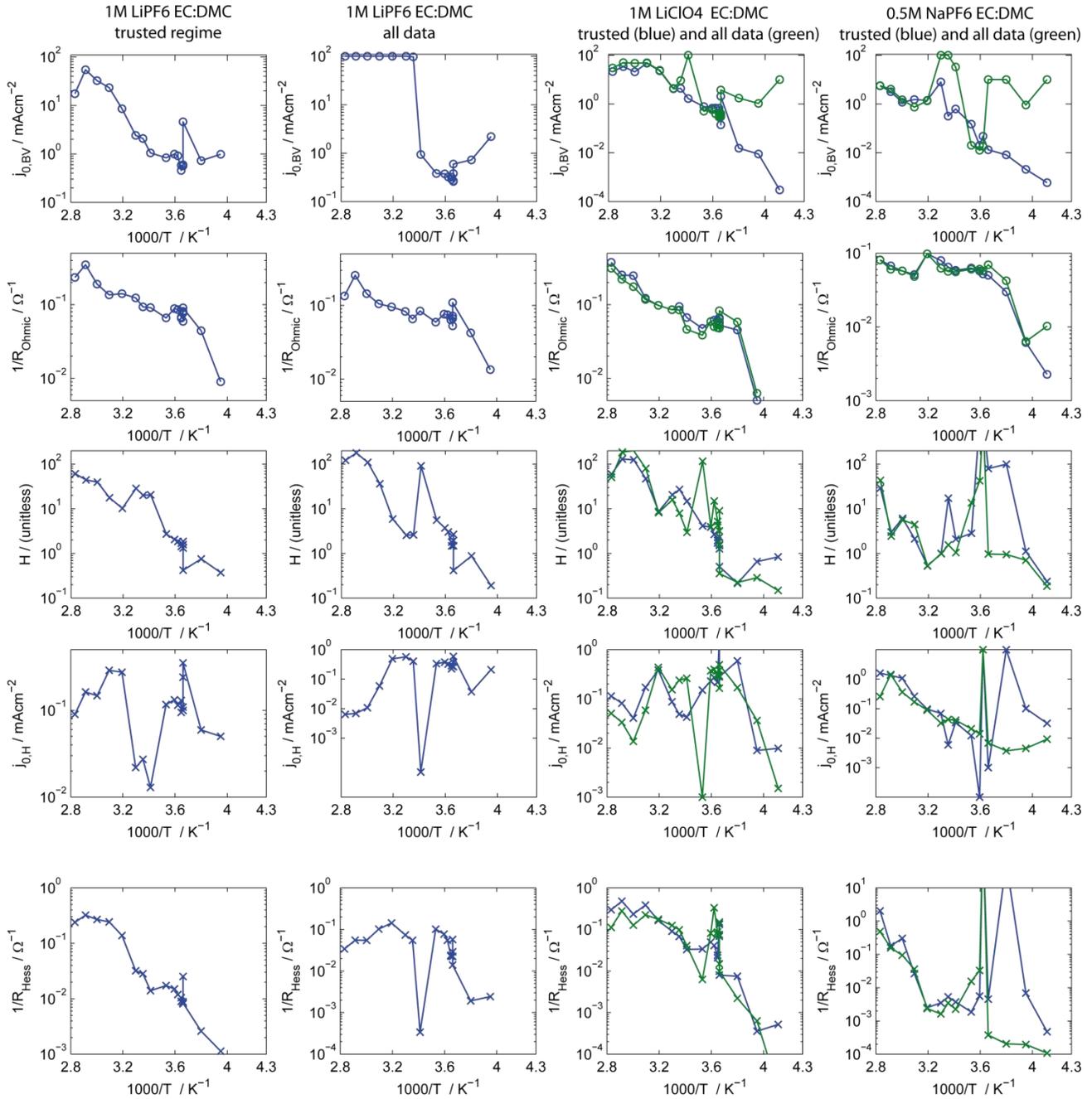

**Supplementary Figure S14: Parameter evolution with standard least-square method:** Matlab's function "lsqcurvefit" applied with equal weighting to either the full data set of extracted initial overpotentials or a trusted region neglecting charge and very high current density discharge cycles for three different electrolyte systems; equal weighting of every data point actually favors values of high current density with high overpotential over fitting at low current density; e.g. a 10 mV variation at 56 mAcm$^{-2}$ is weighted similar to the 10 mV variation at 0.018 mAcm$^{-2}$; this leads to huge errors in the Hessian overpotential as it is almost exclusively sensitive to low and very low current densities; the parameters of $j_{0,BV}$ and $R_{Ohm}$ are still good to fit as they are sensitive at high current density, however, the parameters $H$ and $j_{0,H}$ fluctuate over orders of magnitude; this is better shown in SI Fig. 15.

This "bad" fitting at low current density is better illustrated for 1M LiClO$_4$ EC:DMC at 40 and 50°C in SI Fig. S15. These to temperatures correspond to the fourth and fifth data point in column three of Fig S14. While both $j_{0,BV}$ and $R_{Ohm}$ have smooth trends from 25-70°C, $H$ and $j_{0,H}$ jump significantly as they do not fit the SEI resistance at very low current density in SI Fig S15 for 40°C but do for 50°C. These two fits have been chosen arbitrarily and do not represent a particular good or bad example. They were chosen by coincidence.

Thus, a method with strong weighting of the low current density was used throughout this study. An exemplary weighting vector is given here but is often adapted for outliers and especially for different temperatures as the very high rates often did not give justified overpotentials at temperature below 0°C. Such an weighting vector is for e.g. 1M LiPF$_6$ EC:DMC GF at 25°C

```
curden_ch = [ -1.858, -1.392, -0.929, -0.696, -0.464,-0.278,-0.185,
-0.0929, -0.0465, -0.0186 ]'./A;
curden_dis =[ 74.4, 55.8, 37.2, 18.6, 9.3, 4.65, 1.86, 1.39, 0.93,
0.697, 0.465, 0.186, 0.093, 0.046, 0.0186]'./A;
overpot_ch = [ … ]';
weights = ([ 0 0.004 0.01 0.04 0.05 0.1 0.2 0.4 0.8 1.6,...
        0 0.001 0.005 0.01 0.1 0.5 1 1.2 2 4 8 16 40 80 200].^0.5)';
```

While this weighting vector is artificial, it allows for quick adaptation of non-trusted values e.g. here the very high rate of 74.4 mA. This artificial weighting vector places a weakness in the evaluation as other weightings might lead to other fitting. However, the extracted parameters in this study are not "the 100% exact" parameter for charge transfer or electrolyte resistance. These parameters are too much influenced by dendrite growth, uncertain asymmetries, macroscopic averaging of microscopic effects, assumption of all current through each individual process without taking side reactions into account and most importantly, only one test cell per electrolyte per temperature except at temperatures near the individual $T_C$'s. Thus, this study should be seen exploratory and not as a study to determine precise parameters for modeling purposes etc.

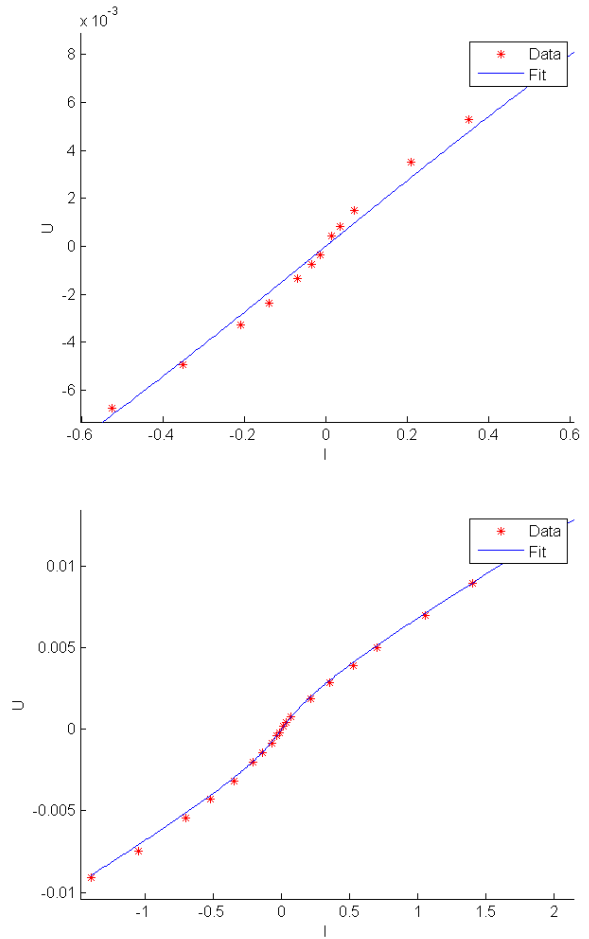

**Supplementary Figure S15: Fitting of standard least-square method with equal weights:** for low and high current density cycles for 1M LiClO$_4$ EC:DMC at a) 40°C and b) 50°C with Matlab's "lsqcurvefit" method corresponding to the data points in SI Fig. S14.

## 6. Suppl. Note 6: Assuming two Hessian resistances

### a) Assuming two SEI transport processes <u>in series</u>

This assumption cannot be distinguished in the current cell setup of two symmetric alkali-electrodes facing each other. Introducing another Hessian resistance in series, therefore, simply increases the number of fitting parameters from initial four to six. Thus a better fitting can be achieved as shown in SI Fig. S16. However, it is not much better than the fit with four parameters in Fig. 3 of MM.

### b) Assuming two SEI transport processes <u>in parallel</u>

Mimicking the transport processes far away from the critical temperature of the SEI transport processes is feasible as one process is significantly higher in resistance. Thus, the less limiting resistances conducts almost all current and the experimental data can be fit neglecting the very limiting transport process that is in parallel. However, both processes conduct near $T_C$ and right at $T_C$ both share the current 50:50. Thus, one can try to model that the assumption $i_{BV} = i_{Ohm} = i_{Hess}$ of the MM changes to the new assumption $i_{BV} = i_{Ohm} = i_{H1} + i_{H2}$ where the associated overpotential through the SEI is the same for both Hessian resistances in parallel $\eta_{H1} = \eta_{H2}$. This method is solved in Matlab with an implicit method since both Hessian resistances and Butler-Volmer are non-linear. The best parameter fit for 1M LiClO$_4$ EC:DMC at 3°C is displayed in SI Fig. S16 with the corresponding fitting parameters listed below:

|  | $j_{0,BV}$ | $R_{Ohm}$ | $H_1$ | $j_{0,H1}$ | $H_2$ | $j_{0,H2}$ |
|---|---|---|---|---|---|---|
| Series | 0.813 | 7.42 | 2.08 | 0.844 | 9.79 | 0.0474 |
| Parallel | 0.406 | 10.34 | 58.15 | 0.0048 | 2.84 | 0.0013 |

Both, series and parallel setting fit the data. But in general, the overall features of the experimental data in j-η plots do not allow precise parameter estimation. Thus further research is necessary to reveal the underlying conduction mechanisms.

In series:

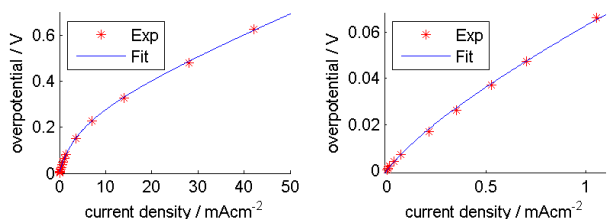

In parallel

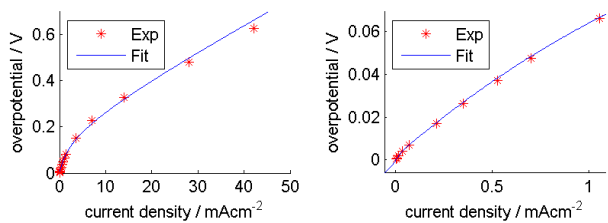

**Supplementary Figure S16: Attempt to fit experiment near $T_C$ with two independent Hessian overpotentials:** 1M LiClO$_4$ in EC:DMC at 3°C modelled with Butler-Volmer, Ohm and two different Hessian either 1$^{st}$ row) in series, 2$^{nd}$ row) in parallel; **in general)** experimental data has too little features for six parameters to be fit instead of the standard four; all processes in series fits good high rates but less good at low rates while two Hessian resistances in parallel in inner SEI sharing current good in fitting low current densities while bad at high current density; but overall not possible to determine parameters correctly.